\documentstyle[a4,12pt]{article}
\begin{document}
\centerline{\bf Criticality in the Integer Quantum Hall Effect}
\vskip 2.0cm
\centerline{Alex Hansen, E.\ H.\ Hauge, Joakim Hove and Frank A.\ Maa{\o}}
\vskip 0.5cm
\centerline{Institutt for fysikk, Norges teknisk-naturvitenskapelige
universitet, NTNU}
\centerline{N--7034 Trondheim, Norway}
\vskip 3.0cm
We review some elementary aspects of the critical properties of the
series of metal-insulator transitions that constitute the integer
quantum Hall effect.  Numerical work has proven essential in charting
out this phenomenon.  Without being complete, we review network
models that seem to capture the essentials of this critical phenomenon.
\footnote{To appear in {\sl Annual Reviews of Computational Physics,\/} 
Vol.\ 5, edited by D.\ Stauffer (World Scientific, Singapore, 1997).}
\vfill
\eject
\bigskip
\centerline{\bf 1.\ Introduction}
\bigskip
\setcounter{footnote}{0}
The existence of localized states in semiconductors has been known since
the seminal work of Anderson [1,2].  He used a deceptively simple-looking 
tight-binding model and showed that when the disorder of the binding energy 
on the scale set by the interaction energy exceeds a given limit (which is
zero in one dimension), the wave functions are localized in space. I.e.\ the
eigenstates fall off exponentially 
while still forming a continuous energy spectrum.  The physics behind this
mechanism is destructive interference.  Another mechanism that leads to
localization is that of percolation, for the first time described
by Flory [3], and later used by 
Skal and Shklovskii [4] to describe hopping resistivity of lightly doped 
semiconductors.  Percolation theory has, however, grown far beyond 
semiconductor physics, and even beyond physics. For a modern
review, see, e.g., Sahimi [5].  The review by 
Thouless [6] discusses percolation and Anderson localization in a unified 
manner.  

In this paper, we shall discuss localization in the integer quantum Hall
effect.  The classical Hall effect is standard textbook material, see
e.g., Ashcroft and Mermin [7]. A conductor in the shape of a thin 
rectangular plate, i.e. a Hall bar, is placed in the strong 
perpendicular magnetic field $\vec B$ as shown in Fig.\ 1. A current $I_x$ 
flows through the bar, giving rise to a longitudinal voltage drop $V_x$ and 
a transversal voltage drop $V_y$, which is also called the Hall voltage.  It 
is the appearance of this transversal voltage that constitutes the classical 
Hall effect.  The ratios
$$
\rho_{xx}={{V_x}\over{I_x}}\;,
\eqno(1)
$$
and
$$
\rho_{yx}={{V_y}\over{I_x}}
\eqno(2)
$$
define the longitudinal resistance and Hall resistance
respectively.\footnote{Note that in two dimensions, resistance and
resistivity are equivalent --- and that this system is essentially
two-dimensional.}  Simple arguments based on the Drude model of conductance
predict the longitudinal resistance to be independent of the magnetic field
$\vec B$, and the Hall resistance to be linear in $B=|\vec B|$.  This
is also found experimentally.
\begin{figure}[htb]
\vspace*{12cm}
\includegraphics{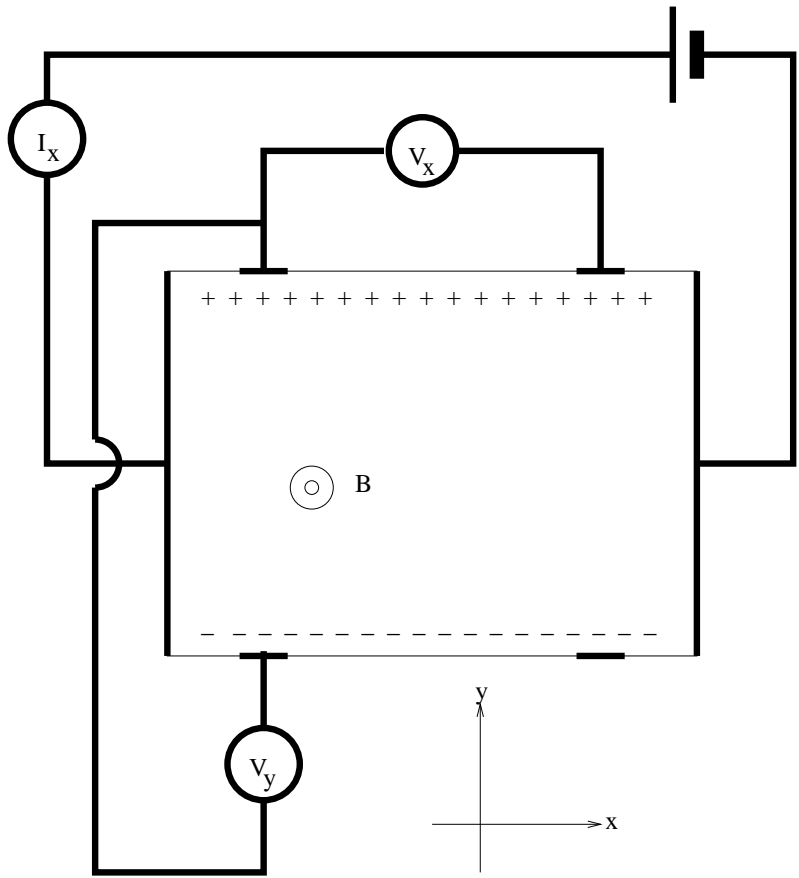}
\caption{Schematic diagram of the Hall system}
\end{figure}

Using a Si MOSFET inversion layer --- which forces the charge carriers
to move in a two-dimensional plane --- placed in a 19 Tesla magnetic field
and at a temperature of 1.5 K, von Klitzing {\it et al.\/} [8] 
observed a completely different behavior of the two resistances:  The 
longitudinal resistance showed a series of pronounced peaks while being zero 
between them.  The Hall resistance did not grow linearly with $B$, but 
developed a series increasingly pronounced steps.  The reader should take 
a look at Paalanen {\it et al.\/} [9] to see these plots, which here are 
based on a GaAs heterostructure (as they did not appear explicitly in the 
original von Klitzing {\it et al.\/} paper). This is the integer quantum 
Hall effect.  We have listed in Table 1 the typical values of some relevant 
physical quantities in the GaAs heterostructure.

The behavior of the longitudinal resistance clearly indicates that the charge
carriers of the quantum Hall system repeatedly localizes and delocalizes
with increasing magnetic field.  What kind of localization goes on, and
why the repeated localization-delocalization-localization transitions?
These questions lie at the heart of the integer quantum Hall effect.

Several very good texts on the integer (and fractional\footnote{Which
appears in clean samples at even higher magnetic fields and lower 
temperatures than are necessary for the integer effect, in the form of steps 
appearing with a different regularity than those of the integer quantum Hall 
effect.}) quantum Hall effect have already appeared, see e.g.\
Ref.\ [10--13]. Our aim with this paper is to convey an elementary 
understanding of the phenomenon, rather than a full-scale review
of the entire literature. We therefore urge the readers to consult
these more complete works.

We should perhaps apologize that the parts of this review that deal with the 
{\it numerical\/} studies of the integer quantum Hall effect
is short compared to those dealing with describing the physics of the 
system.  However, we believe this to be justified.  The numerical models 
are simple,  but the results that they lead to are profound.  In order for
the reader to appreciate the numerical models fully, we have chosen to
emphasize the physics of the effect.  We write this review for an audience
in mind with a background in statistical 
\vskip1.0cm 
\centerline{
\vbox{\offinterlineskip
\hrule
\halign{&\vrule#&
\strut\quad\hfil#\quad\cr
\noalign{\hrule}
height2pt&\omit&&\omit&\cr
&$m=0.067m_e$\qquad&&Effective mass\qquad&\cr
&$n=4\cdot 10^{15}{\rm m}^{-2}=(16{\rm nm})^{-2}$\qquad&
&Electron sheet density\qquad&\cr
&$E_F=14$meV\qquad&&Fermi energy\qquad&\cr
&$\lambda_F=40$nm\qquad&&Fermi wavelength\qquad&\cr
&$\Lambda=10^2$ -- $10^4$nm\qquad&&Elastic mean free path\qquad&\cr
&$\Lambda_\phi > \Lambda$\qquad&&Phase coherence length\qquad&\cr
&$l=\sqrt{\hbar/eB}=26(B/{\rm T})^{-1/2}$nm\qquad      
&&Magnetic length\qquad&\cr
height2pt&\omit&&\omit&\cr}
\hrule}
}
\vskip1.0cm
\centerline{Table 1: Some relevant physical quantities in the GaAs heterostructure}
\vspace*{1.5cm}
physics.  We have therefore 
attempted to be quite thorough and explicit in setting up the physical
framework. No prior knowledge to the problem of two-dimensional quantum
transport is assumed, and only elementary quantum mechanics is used.  The 
essence of what will follow is this:  The integer quantum Hall effect 
is closely related to the Landau levels of the electrons, formed
when the system is placed in a strong perpendicular magnetic field.  The 
energy of the electrons is quantized in way similar to the harmonic 
oscillator, see Eqs.\ (13) and (15).  Whenever the magnetic field is reduced
and one of these Landau levels in the bulk sinks through the Fermi level 
of the electrons, an additional contribution to the electrical conductance is 
made possible by filling this new Landau level with electrons from the 
attached reservoirs. This leads to the quantized current of Eq.\ (68) and to 
the subsequent ``hydrodynamical" scenario thereafter in which disorder plays 
a crucial r{\^o}le.

The organization of the review is as follows: We discuss in the next section 
the behavior of a
two-dimensional free electron gas in a strong perpendicular magnetic field.
The integer quantum Hall effect enters in a very subtle way in this 
simple model.  When we introduce interactions between the underlying lattice 
and the electrons, the integer quantum Hall effect becomes much more 
robust --- and easy to understand. We discuss this in Sec.\ 3.  As 
will be evident from this section, percolation plays an important r{\^o}le. 
In Sec.\ 4, we discuss the critical behavior of the integer quantum Hall
system, which is caused by localization-delocalization transitions when the
the externally applied magnetic field is changed.  We furthermore discuss
a famous semi-classical argument that seemingly identifies the nature and
thus the universality class of these transitions.  Unfortunately, there are
problems with this argument, as we point out. In Sec.\ 5, we discuss 
a class of numerical network models that seems to capture the critical
behavior of the quantum Hall system very well.  Section 6.\ contains our
conclusions.


\newpage
\bigskip
\centerline{\bf 2.\ The Free Electron Gas in a Closed System}
\bigskip
The Hamiltonian of an electron moving in the $(x,y)$ plane and in a
perpendicular, constant magnetic field $\vec B=B\vec e_z$ is
$$
H_0={1\over {2m}}\left(\vec p+e\vec A \right)^2\;,
\eqno(3)
$$
where $\vec B=\vec\nabla\times A$, $e$ is the elementary charge and $m$
the effective mass.
We shall temporarily not specify any particular gauge.  The standard 
commutation relations apply:
$$
[x,p_x]=[y,p_y]=i\hbar\;,
\eqno(4)
$$
and
$$
[x,y]=[p_x,p_y]=[x,p_y]=[y,p_x]=0\;.
\eqno(5)
$$
Following Kubo {\it et al.\/} [14], we perform a canonical transformation 
from the original coordinates $(x,y)$ and $(p_x,p_y)$ to the relative 
coordinates $(\xi,\eta)$ and guiding center coordinates $(X,Y)$:
$$
\xi={1\over{eB}}\left(p_y+eA_y\right)\;,
\eqno(6)
$$
$$
\eta=-{1\over{eB}}\left(p_x+eA_x\right)\;,
\eqno(7)
$$
$$
X=x-\xi\;,
\eqno(8)
$$
and
$$
Y=y-\eta\;.
\eqno(9)
$$
The commutation relations (4) and (5) lead to
$$
[X,Y]=-[\xi,\eta]=i\left({{\hbar}\over{eB}}\right)=il^2\;,
\eqno(10)
$$
and
$$
[\xi,X]=[\eta,Y]=[\xi,Y]=[\eta,X]=0\;.
\eqno(11)
$$
In Eq.\ (10), we have used the magnetic length
$$
l^2={\hbar\over{eB}}\;,
\eqno(12)
$$
which corresponds to the cyclotron radius for the lowest quantum
state.  The accompanying classical cyclotron frequency is
$$
\omega={{eB}\over{m}}\;.
\eqno(13)
$$
The relative coordinates $\xi$ and $\eta$ describe the position of the
electron relative to the center of the circle on which the electron is
moving.  The position of this center, the guiding center, is given by
$X$ and $Y$. The Hamiltonian is in these coordinates
$$
H_0={m\over{2}}\omega^2\left(\xi^2+\eta^2\right)\;,
\eqno(14)
$$
which we recognize as the harmonic oscillator Hamiltonian by appealing
to the commutation relation (10).  The energy spectrum is therefore
$$
E_n=\hbar\omega \left( n+{1\over 2}\right)\;.
\eqno(15)
$$
These are the {\it Landau levels.\/}

In order to construct the wave functions, we must specify a gauge.  We
choose the Landau gauge where $\vec A^x = (-By,0)$.  The Schr{\"o}dinger
equation then becomes
$$
-{{\hbar^2}\over{2m}}\left[{{\partial^2}\over{\partial y^2}}+
\left({{\partial}\over{\partial x}}-{{ieB}\over{\hbar}} y\right)^2
\right]\psi_n^x=E_n\psi_n^x\;,
\eqno(16)
$$
where the superscript $x$ refers to the choice of gauge that we have made.
We now set
$$
\psi_{n,Y}^x(x,y)={{e^{ikx}}\over{\sqrt{L}}}\tilde\chi_n(y)\;,
\eqno(17)
$$
so that Eq.\ (16) becomes
$$
\left[-{{\hbar^2}\over{2m}}{{d^2}\over{d y^2}}
+{{m}\over 2}\omega^2\left(y-kl^2\right)^2 \right] \tilde\chi_n(y)
=E_n \tilde\chi(y)\;.
\eqno(18)
$$
Thus, the energy eigenfunctions $\psi_{n,Y}^x$ consist of plane waves in the 
positive $x$ direction and harmonic oscillator wave functions 
$\tilde\chi_n(y)=\chi_n(y-kl^2)$ in the $y$ direction centered around 
$Y=kl^2$.  The energy levels are independent of $k$.  The energy 
eigenfunctions are therefore
$$
\psi^x_{n,Y}(x,y)={{e^{iY x/l^2}}\over{\sqrt{L}}}\chi_n(y-Y)\;.
\eqno(19)
$$
These eigenfunctions form a complete set, and any wave function may be
written
$$
\psi^x(x,y)=\sum_n\int dY {{e^{iY x/l^2}}\over{\sqrt{L}}}
\chi_n(y-Y) C_n(Y)\;.
\eqno(20)
$$

A change of gauge $\vec A \to \vec A -\vec\nabla \alpha$ leads to a change
in phase of the wave function, $\psi\to\exp(ie\alpha/\hbar)\psi$.  For
example, we may go from Landau gauge $\vec A^x=(-By,0)$ to the symmetric 
gauge $\vec A^s=(-By/2,Bx/2)$.  Integrating the difference between these,
$\vec\nabla\alpha=$$(-By/2,-Bx/2)$, gives $\alpha=-Bxy/2$.  An arbitrary wave 
function in this gauge may therefore be written
$$
\psi^s(x,y)=\sum_n\int dY {{e^{-i(xy-2Yx)/2l^2}}\over{\sqrt{L}}}
\chi_n(y-Y) C_n(Y)\;.
\eqno(21)
$$

Let us now count the number of states per Landau level and per area in the
system.  Assuming periodic boundary conditions in the
$x$ direction, the wave function (17) must be periodic in the length
$L$ of the system, $\psi_{n,Y}^x(x+L,y)=\psi_{n,Y}^x(x,y)$.
This quantizes the possible values of $Y$,
$$
Y={{2\pi l^2}\over L} j\;,
\eqno(22)
$$
where $j=0,\pm 1,\pm 2,\cdots$.  The system has width $W$ in the $y$
direction. Thus, $0\le Y\le W$ --- remember that $Y$ determines the center
of the harmonic oscillator wave functions in the $y$ 
direction.\footnote{We ask for patience from those who feel that our 
treatment of the boundaries  of the system in the $y$ direction to be too 
simple minded.  We will return to these boundaries in great detail in
Sec.\ 3.2.} Combination of this restriction with Eq.\ (22), leads to $j$ 
being confined to the integers on the interval 
$0\le j \le LW/2\pi l^2=j_{\max}$.  Thus, the number of states per Landau 
level per area is
$$
n_B={{j_{\max}}\over{LW}}={1\over{2\pi l^2}}\;.
\eqno(23)
$$

We now turn on an electric field $\cal E$ in the $y$ direction.  The
Hamiltonian (14) then becomes
$$
H={m\over 2}\omega^2\left(\xi^2+\eta^2\right)+
e{\cal E}_y\left(Y+\eta\right)\;,
\eqno(24)
$$
where we have used Eq.\ (9), $y=Y+\eta$.  We may rewrite this Hamiltonian
as
$$
H={m\over 2}\omega^2\left(\xi^2+\left(\eta
+{{e{\cal E}_y}\over{m\omega^2}}\right)^2\right)
-{{e^2{\cal E}_y^2}\over{2m\omega^2}}+e{\cal E}_yY\;.
\eqno(25)
$$
From this it is possible to write down the eigen\-functions and corresponding
energy levels directly by inspection:
$$
\psi^x_{n,Y}(x,y)={{e^{iY x/l^2}}\over{\sqrt{L}}}
\chi_n\left(y-Y+{{e{\cal E}_y}\over{m\omega^2}}\right)\;,
\eqno(26)
$$
and
$$
E_{n,Y}=\hbar\omega \left( n+{1\over 2}\right)
+e{\cal E}_y\left(Y-{{e{\cal E}_y}\over{m\omega^2}}\right)+{m\over 2}
\left({{{\cal E}_y}\over{B}}\right)^2\;,
\eqno(27)
$$
where the second term represents potential energy 
in the electric field and the third term the kinetic energy associated with 
guiding-center drift.

Let us now calculate the current densities in the 
system.\footnote{By using the expression
$$
v ={1\over\hbar}{{\partial E_{n,Y}}\over{\partial k}}\;,
$$
where $E_{n,Y}$ is given by Eq.\ (27), we obtain the results Eqs.\ (34) 
and (35) immediately.  However, the way we present the derivation
of Eqs.\ (34) and (35) is more detailed, and is in fact a 
derivation of the above equation.}  Essential ingredients of such a 
calculation are the expectation values of the velocity in the $x$ and $y$ 
directions,
$$
v_x(n,Y)=\int dx\ dy\  {\psi^x_{n,Y}}^*(x,y) \dot x \psi^x_{n,Y}(x,y)\;,
\eqno(28)
$$
and
$$
v_y(n,Y)=\int dx\ dy\  {\psi^x_{n,Y}}^*(x,y) \dot y \psi^x_{n,Y}(x,y)\;.
\eqno(29)
$$
We split $\dot x$ and $\dot y$ into $\dot x=\dot \xi+\dot X$ and
$\dot y =\dot \eta + \dot Y$.  The Heisenberg equations of motion give
$$
{{d\xi}\over{dt}} ={{i}\over{\hbar}}[H,\xi]=-\omega\left(\eta+
{{e{\cal E}_y}\over{m\omega^2}}\right)\;,
\eqno(30)
$$
$$
{{d\eta\ }\over{dt}} ={{i}\over{\hbar}}[H,\eta]=+\omega\xi\;,
\eqno(31)
$$
$$
{{dX}\over{dt}}={{i}\over{\hbar}}[H,X]={{el^2}\over{\hbar}}\ {\cal E}_y
={{{\cal E}_y}\over B}\;,
\eqno(32)
$$
and
$$
{{dY}\over{dt}} ={{i}\over{\hbar}}[H,Y]=0\;.
\eqno(33)
$$
We have here used the commutation relations Eqs.\ (10) and (11),
and the Hamiltonian, Eq.\ (24).  Combining these results with equations
(28) and (29) give
$$
v_x(n,Y)={{{\cal E}_y}\over{B}}\;,
\eqno(34)
$$
and
$$
v_y(n,Y)=0\;.
\eqno(35)
$$
In Eq.\ (34) we have used that 
$$
\int_{-\infty}^\infty d\eta\  
|\chi_n(\eta+e{\cal E}_y/m\omega^2)|^2 (\eta+e{\cal E}_y/m\omega^2)
=\int_{-\infty}^\infty d\eta\  |\chi_n(\eta)|^2\eta=0\,.  
\eqno(36)
$$
In equation 
(35), we have used that 
\setcounter{equation}{36}
\begin{eqnarray}
\int_{-\infty}^\infty d\eta\ 
\chi_n(\eta+e{\cal E}_y/m\omega^2)^*
\xi\chi_n(\eta+e{\cal E}_y/m\omega^2) & = & 
\nonumber \\
\int_{-\infty}^\infty d\eta\ 
\chi_n(\eta+e{\cal E}_y/m\omega^2)^*(-il^2){{d}\over{d\eta\ }}
\chi_n(\eta+e{\cal E}_y/m\omega^2) & = & 0\;.
\end{eqnarray}

With $n_e$ electrons per area, Eqs.\ (34) and (35) give rise to a current 
density in the $x$ direction equal to
$$
j_x=-\sum_{n,Y}ev_x(n,Y)
=-2\pi l^2 {n_e}{{e^2}\over{h}}\ {\cal E}_y
=-{{n_e}\over{n_B}}\ {{e^2}\over{h}}\
{\cal E}_y=-{{e^2\nu}\over{h}}\ {\cal E}_y\;.
\eqno(38)
$$
The sum runs over all electrons and we ignore spin. In the $y$ direction the 
current density is
$$
j_y=-\sum_{n,Y}ev_y(n,Y)=0\;.
\eqno(39)
$$
We have in Eq.\ (38) defined the {\it filling factor\/}
$$
\nu={{n_e}\over{n_B}}\;.
\eqno(40)
$$
$n_B$ is defined in Eq.\ (23).  

From Eqs.\ (38) and (39), we read off the elements 
of the conductivity 
$$
\sigma_{xx}=\sigma_{yy}=0\;,
\eqno(41)
$$
and
$$
\sigma_{yx}=-\sigma_{xy}={{e^2}\over{h}}\ \nu\;.
\eqno(42)
$$
From this the resistance tensor follows by inversion, or directly by
reference to Eqs.\ (1) or (2) as
$$
\rho_{xx}=\rho_{yy}=0\;.
\eqno(43)
$$
and 
$$
\rho_{xy}=-\rho_{yx}={{h}\over{e^2\nu}}\;.
\eqno(44)
$$

The resistances (43) and (44) capture one aspect of what is measured in the 
integer quantum Hall effect: zero longitudinal resistance.  
However, the steps, 
which have their centers at {\it integer values of the filling factor\/} 
$\nu$ and the onset of longitudinal conductance when the system moves from 
one step to the next, are not found in this system.  The reason that the 
steps are not visible, is that the system is closed.  In Sec.\  3.2, we will
consider an open system, where steps do appear.  The reason why steps only
appear in open systems will be clear at that point.  Before we turn to
this discussion, we add interactions between the electrons and the
underlying system, as they play a crucial r{\^o}le in what happens 
between the steps --- which is the subject of this review.
  
\bigskip
\centerline{\bf 3.\ Interactions between Electrons and Lattice: The
Disordered Potential}
\bigskip
We now add an arbitrary potential $V(x,y)$ to the Hamiltonian (3).  
The Schr{\"o}\-dinger equation we have to solve is then
$$
H\psi(x,y)=H_0\psi(x,y)+V(x,y)\psi(x,y)=E\psi(x,y)\;.
\eqno(45)
$$
We expand the wave function as in Eq.\ (20)  In this basis, we have
that
$$
H_0\psi^x(x,y)=\sum_{n'}\hbar\omega\left(n'+{1\over 2}\right)
\int dY'\ e^{iY' x/l^2}\chi_{n'}(y-Y') C_{n'}(Y')/{\sqrt{L}}\;.
\eqno(46)
$$
By multiplying Eq.\ (45) by $e^{-iYx/l^2} \chi^*_{n}(y-Y)/\sqrt{L}$
and integrating over $(x,y)$, the Schr{\"o\-}dinger equation is transformed
into
\setcounter{equation}{46}
\begin{eqnarray}
\int dx\ dy\  V(x,y) \sum_{n'}\int dY'\ e^{i(Y'-Y)x/l^2}
\chi^*_n(y-Y)\chi_{n'}(y-Y') C_{n'}(Y')/2\pi l^2 & &
\nonumber \\
=(E-E_n)C_n(Y) \;. & &
\end{eqnarray}
This expression is similar, but not identical, to the one given 
by Tsukada [15], which is the standard reference, but contains a
small error.

Let us now make some changes of variables in Eq.\ (47). By defining
$\Delta_Y=Y'-Y$ and using that $x=X+\xi$ and $y=Y+\eta$, 
we can write Eq.\ (47) as
\begin{eqnarray}
\int d\xi d\eta\  V(X+\xi,Y+\eta)\int d\Delta_Y 
{{e^{i\Delta_Y(X+\xi)/l^2}}\over{2\pi l^2}} & &
\nonumber \\
\sum_{n'} \chi^*_n(\eta)\chi_{n'}(\eta-\Delta_Y) C_{n'}(Y+\Delta_Y)
& = & (E-E_n)C_n(Y)\;.
\end{eqnarray}

So far, we have made no approximations.  However, from now on we shall assume
that $V(x,y)$ is slowly varying on the scale of the magnetic length $l$.
Since the functions $\chi_n(x)$ are localized over distances of the order of
$l$ --- they are harmonic oscillator eigen\-functions --- we may substitute
$V(x,y)$ by $V(X,Y)$ in Eq.\ (48) as the lowest order term in
an expansion in $l$.  With this substitution, Eq.\ (48) simplifies
dramatically by first integrating over $\xi$ and then using the 
orthonormality of the basis functions $\chi_n$,
$$
V(X,Y)C_{n}(Y)=(E-E_n)C_n(Y)\;.
\eqno(49)
$$
The corresponding energy eigenfunctions are 
$$
\psi^x_n(x,y)=\int dY {{e^{iY x/l^2}}\over{\sqrt{L}}}
\chi_n(y-Y) C_n(Y)\;.
\eqno(50)
$$

What is the physical contents of Eq.\ (49)?  
In the classical limit, where $X$ and $Y$ are treated as c-numbers,
it predicts a motion that follows the equipotential curves of $V(x,y)=E-E_n$.
The classical electrons move in a strong magnetic field such that they neither
gain nor loose energy. In order to expand on this, we write down the 
Heisenberg equations of motion for $X$ and $Y$:
$$
{{dX}\over{dt}}={{i}\over{\hbar}}[H,X]
=+{1\over{eB}}\ {{\partial V}\over{\partial y}}\;,
\eqno(51)
$$
and
$$
{{dY}\over{dt}}={i\over{\hbar}}[H,Y]
=-{1\over{eB}}\ {{\partial V}\over{\partial x}}\;,
\eqno(52)
$$
where $H=H_0+V$, where $H_0$ is given in Eq.\ (3), and we have used
the commutation relations (10) and (11).\footnote{Write
$V$ as a Taylor series in $x$ and $y$, then calculate the commutators
term by term and reassemble the result.}  
There are no approximations in Eqs.\ 
(51) and (52). Introducing the same level of approximation as was done to 
arrive at Eq.\ (49), i.e., assuming that $V$ is
slowly varying on the scale of the magnetic length $l$, Eqs.\ (51)
and (52) become\footnote{If the potential $V$ is not caused by
the surrounding medium in which the electron move, but rather their mutual
interactions, Eqs.\ (53) and (54) become in the classical limit (when $X$ 
and $Y$ are c-numbers) the Kirchhoff equations of motion for vortices.  These 
have many strange properties, for example that the three-body problem is 
integrable, see Gr{\"o}bli [16], Synge [17] and Aref [18].}
$$
{{dX}\over{dt}}=+{1\over{eB}}\ {{\partial V}\over{\partial Y}}\;,
\eqno(53)
$$
and
$$
{{dY}\over{dt}}=-{1\over{eB}}\ {{\partial V}\over{\partial X}}\;.
\eqno(54)
$$
These equations are also valid in the classical limit.  Let us now
multiply the velocity vector of the guiding centers $(\dot X, \dot Y)$
with the gradient of the potential, $\vec\nabla V=(\partial V/\partial X,
\partial V/\partial Y)$.  We then get $\dot{\vec U}\cdot\vec\nabla V=0$
by use of Eqs.\ (53) and (54).  Here $\vec U = (X,Y)$.
Thus, the motion of the guiding center is 
always orthogonal to the local gradient of the interaction potential, 
and as a consequence, the guiding centers follow the equipotential curves 
of $V$.

\bigskip
\centerline{\bf 3.1 Motion Near a Saddle Point: Tunneling}
\bigskip
There will be quantum mechanical 
tunneling if two distinct equipotential lines come close at
some point in space, as shown in Fig.\ 2.  This may be somewhat
surprising in that the tunneling has to appear in the direction
{\it orthogonal\/} to the paths.  We may understand this qualitatively
through the following semi-classical picture: The paths drawn as in Fig.\
2 are those of the guiding centers, which the electron itself is
rotating about.  Thus, the momentum vector of the electron points in ``all"
directions, even orthogonal to the path that the guiding center is
following --- making tunneling possible in the seemingly most unlikely
direction.
\begin{figure}[h]
\vspace*{3.4cm}
\includegraphics{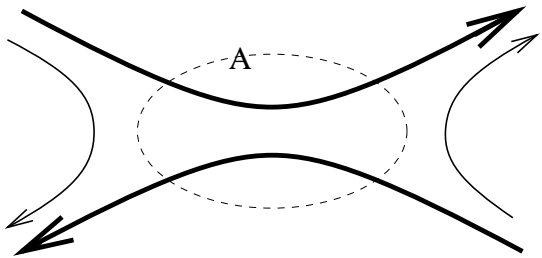}
\caption{The potential $V$ has a saddle point in
region $A$.  Even if the two paths marked in bold, which follow
the equipotential curves of $V$, are antiparallel where they are at
the closest, there is tunneling between them.}
\end{figure}

Let us analyse this transverse tunneling more closely.  Following Fertig
[19] and Mil'nikov and Sokolov [20], we concentrate on the region near a 
saddle point of $V$. We choose a coordinate system so that the saddle point 
is placed at the origin.  Furthermore, we orient the coordinate system so 
that close to the origin, we have that
$$
V(X,Y)=V_0-\alpha X^2+\beta Y^2\;,
\eqno(55)
$$
where $\alpha>0$ and $\beta >0$.  Thus, $V$ has a local maximum at the origin
when moving along the $X$ axis, and a local minimum at the origin when moving
along the $Y$ axis.  Two situations may then occur:  (1) $E-E_n<V(0,0)=V_0$,
where $E$ is found by solving the Schr{\"o}dinger equation (49).  The
corresponding semi-classical paths are shown in Fig.\ 3a.  Likewise, the 
situation when $E-E_n>V_0$ is shown in Fig.\ 3b. Suppose that
the situation is that of Fig.\ 3a.   Then, it is convenient to use the 
representation $Y=il^2d/dX$, thus satisfying the commutator $[X,Y]=il^2$, 
Eq.\  (10). Equation (49) then becomes\footnote{Since the commutator 
is of ${\cal O}(l^2)$, one could worry about the consistency of Eq.\ 
(56). A tedious check leads to a reassuring result: The only extra term 
needed for consistency represents a constant shift in energy near the saddle 
point.  We ignore this irrelevant constant in what follows.}
$$
\left(-\beta l^4{{d^2}\over{dX^2}}-\alpha X^2\right)C_n(X)
=\left(E-E_n-V_0\right)C_n(X)\;.
\eqno(56)
$$
If, on the other hand, the situation is the one described in Fig.\ 3b,
we switch to the representation $X=-il^2d/dY$, which also satisfies the
commutation rule $[X,Y]=il^2$.  Equation (49) then takes the form,
after an overall change of sign,
$$
\left(-\alpha l^4{{d^2}\over{dY^2}}-\beta Y^2\right)C_n(Y)
=\left(V_0+E_n-E\right)C_n(Y)\;.
\eqno(57)
$$
\begin{figure}[h]
\vspace*{5.5cm}
\includegraphics{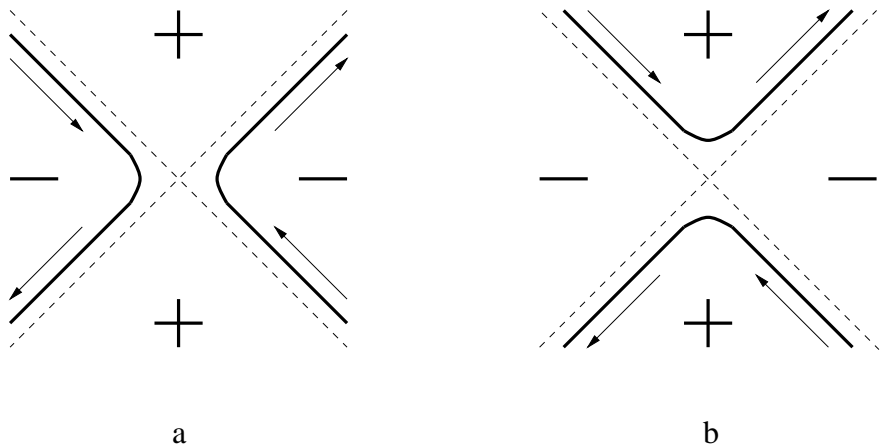}
\caption{The potential $V$ has a saddle point at the
origin. In (a) we show two semi-classical paths corresponding to $E<E_n+V_0$,
and in (b) we show the paths corresponding to $E>E_n+V_0$.}
\end{figure}

Thus, we are faced with solving the one-dimensional scattering problems
posed by the Schr{\"o}dinger Eqs.\ (56) and (57) in order to study tunneling.  
This is difficult since $|V|\to\infty$ when $|X,Y|\to \infty$. However, 
we can get around this difficulty by substituting 
$V_0-\alpha X^2\to V_0/\cosh^2(\sqrt{\alpha/V_0} X)$ in Eq.\ (56)
or $-V_0+\beta Y^2\to V_0/\cosh^2(\sqrt{\beta/V_0} Y)$ in Eq.\ (57),
see Hansen [21].  This substitution leads to the correct quadratic behavior 
near the origin, and plane-wave solutions far away from the origin.  The
exact solution of the resulting one-dimensional scattering problem
may be found, {\it e.g.,\/} in ter Haar [22] or Landau and Lifshitz [23].  
Both Eqs.\ (56) and (57) lead to a transmission coefficient 
$$
T={{P}\over{1+P}}\;,
\eqno(58)
$$
where
$$
P=e^{-2\pi/l^2 \sqrt{1/\alpha\beta} |E-E_n-V_0|}\;.
\eqno(59)
$$
We make a couple of comments: (1) When $\Delta E_n=(E-E_n-V_0)\to 0$, the
transmission coefficient $T$ approaches the value 1/2, which is the maximum
value it may take.  Thus, when $\Delta E_n=0$, an electron approaching
the saddle point will leave it following either of the two possible paths
with equal probability. (2) Using the WKB approximation, as was the approach 
in Mil'nikov and Sokolov [20], leads to $T_{WKB}=P=T/(1-T)=T+T^2+\cdots$.

The potential $V(x,y)$ is caused by disorder in or near the inversion
layer of the MOSFET, which in turn may be induced by, {\it e.g.,\/} crystal
defects or impurities.  Thus, the potential may be seen as a random
landscape of valleys and ridges.  Electrons move along equipotential
curves in this landscape.  Most of these curves form closed loops, which may
join across saddle points.  How can such a scenario give rise to the
strange behavior of the conductivity that one observes in the integer quantum
Hall effect?  Before attempting to answer this question, we discuss the
concept of conductivity in the present context.

\bigskip
\centerline{\bf 3.2 Conductivity and Edge States}
\bigskip
Conductivity is a local quantity.  Conductance is not.  In the following
discussion, it will appear that conductance is the fundamental quantity
in the quantum Hall system, rather than the conductivity.  
However, this may be misleading, and just a result of the framework we 
use to present the results.  To make an analogy, presenting the Maxwell 
equations in the form of surface and line integrals makes them look very 
different, but the physics is of course the same.  We follow in this section 
mainly the work of B{\"u}ttiker [24]. 

So far, we have not cared about the edges in our Hall system of size
$L\times W$.  The only place where they have entered the discussion,
is where we counted the number of states per area.  Let us look at
the solutions of the Schr{\"o}dinger Eq.\ (45) for the
Hall system confined to a bar of finite size, see Halperin [25].  First,
we assume that there is no disorder potential $V$.  We have already
determined the energy eigenfunctions for the infinite system, Eq.\
(19).  These eigenfunctions, $\psi^x_{n,Y}(x,y)$, consist of
``stripes" oriented in the $x$ direction of width $\sim l$.  The position
of the middle of a given ``stripe" is determined by the quantum number $Y$.
The wave functions must have nodes at $y=0$ and $y=W$, i.e., at the
edges of the Hall bar.  Thus, for $Y=0$ or $Y=W$, 
only the odd harmonic oscillator
wave functions, $\chi_{2m+1}(y)$ fulfil the boundary condition.  Thus,
$2m+1=1\Rightarrow m=0$ 
corresponds to the lowest energy eigenstate here, and the
corresponding energy level is $E_{0,Y=0}=3\hbar\omega/2$.  Moving away from
the edge and into the sample, i.e., increasing $Y$, the energy
levels move continuously to their bulk values, which for the ground state
is $E_{0,Y}=\hbar\omega/2$.  This is general:  The energy eigenvalue
corresponding to the quantum number $n$ takes on the value $E_{n,Y}=
\hbar\omega(n+1/2)$ inside the sample, but rises sharply to the values
$E_{n,Y=0}=E_{n,Y=W}=\hbar\omega(2n+3/2)$ at the edges.  This is illustrated
in Fig.\ 4.  It should be noted in this figure that the energy 
levels $E_{n,Y}$ is continued for $Y$ values that are less than 0 or larger
than $W$.  This makes sense.  Classically, this means only that the 
{\it center\/} of the semicircles that the electrons follow lie outside the 
sample.  The charges themselves, whose coordinates are $(x,y)$, never move 
outside the sample limited by $y=0$ and $y=W$.
\begin{figure}[h]
\vspace*{11cm}
\includegraphics{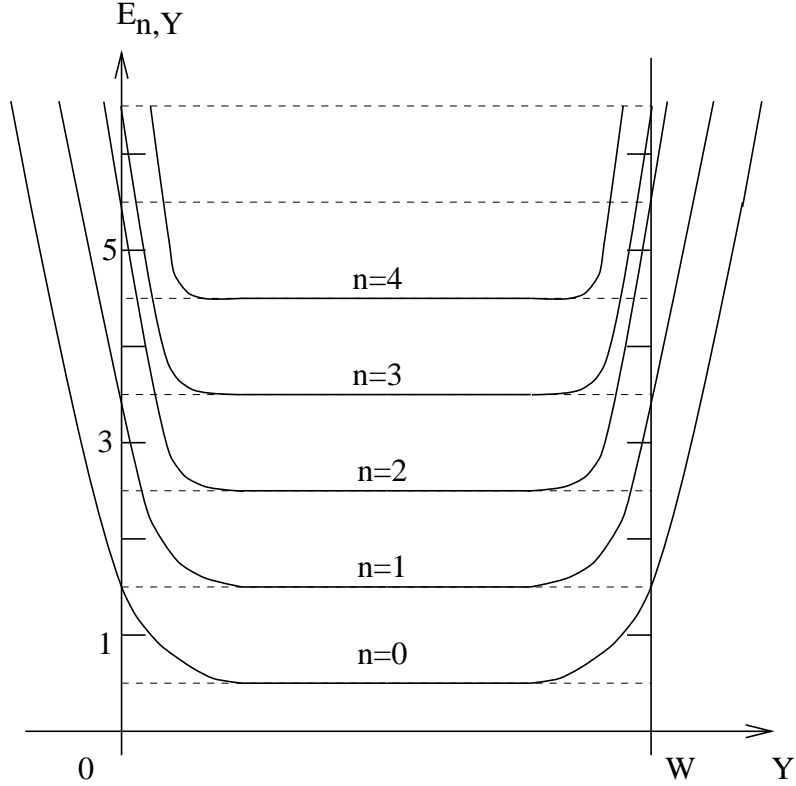}
\caption{The energy eigenvalues $E_{n,Y}$ for the free electron gas in 
a sample of width $W$.}
\end{figure}

Are there any currents flowing in this system?  Judging from Eqs.\ (38)
and (39), the answer is no when there is no external electric
field ${\cal E}_y$ present.  However, this is not true, as we did not 
analyse what is happening at the edges of the sample.  From Eqs.\ 
(28) and (36), we have that\footnote{We are implicitly
assuming periodic boundary conditions in the $x$ direction by our choice
of wave functions, Eq.\ (19).  Thus, we only discuss the boundaries
$y=0$ and $y=W$.  This is known in the literature as {\it Corbino ring\/}
geometry.}
$$
v_x(n,Y)=-\omega\int_0^W d y |\tilde\chi_{n,Y}(y-Y)|^2(y-Y)\;,
\eqno(60)
$$
where $\tilde\chi_{n,Y}(y)\to\chi_{2n+1}(y)$ when $Y\to0$ or $W$, and
$\tilde\chi_{n,Y}(y)\to\chi_n(y)$ when $Y\gg0$ and $Y\ll W$.
The wave function $\tilde\chi_n$ 
falls off over a range $l$ on each side of $Y$,
except at the borders, where it is zero outside the Hall bar.  Thus, when
$Y\gg 0$ and $Y\ll W$, Eq.\ (60) integrates to zero --- as was already
used in Eq.\ (36).  However, when $Y=0$, Eq.\ (60) leads to
$$
v_x(n,0)=-\omega\int_0^\infty dy\ |\chi_{2n+1}(y)|^2 y < 0\;,
\eqno(61)
$$
and for $Y=W$,
$$
v_x(n,W)
=-\omega\int_{-\infty}^W dy\ |\chi_{2n+1}(y-W)|^2 (y-W)=-v_x(n,0)\;.
\eqno(62)
$$
There are therefore currents flowing in opposite directions along the edges 
of the system.  

Let us now attempt to rederive the conductances, Eqs.\ (41) and (42), but 
{\it without\/} applying an external field ${\cal E}_y$ explicitly.  The 
reason for this is that it is awkward to work with small but finite external 
fields when the disorder potential $V(x,y)$ is present.  And the reason why 
one should believe that to be possible, is that conductance is an equilibrium 
property of the system. This will be clear in a moment.  We shall find 
that we fail slightly, but in an interesting way.

Let us calculate
\setcounter{equation}{62}
\begin{eqnarray}
{{dE_{n,Y}}\over{dY}}& = & \int_0^W dy\ \tilde\chi_{n,Y}^*(y-Y)
{{dE_{n,Y}}\over{dY}}\tilde\chi_{n,Y}(y-Y) 
\nonumber \\
& = & \int_0^W dy\ \tilde\chi_{n,Y}^*(y-Y){{dH_0}\over{dY}}
\tilde\chi_{n,Y}(y-Y) 
\nonumber \\
& = & {{m\omega^2}\over 2}\int_0^W dy\ \tilde\chi_{n,Y}^*(y-Y){d\over{dY}}
\left[\xi^2+\left(y-Y\right))^2\right]
\tilde\chi_{n,Y}(y-Y) 
\nonumber \\
& = & -m\omega^2\int_0^W dy\ \tilde\chi_{n,Y}^*(y-Y)\left[y-Y\right]
\tilde\chi_{n,Y}(y-Y)\;.
\end{eqnarray}
Finally, by combining this expression with Eq.\ (61), we derive the 
dispersion relation
$$
v_x(n,Y)={{l^2}\over{\hbar}}\ {{dE_{n,Y}}\over{dY}}\;.
\eqno(64)
$$
Once again, we see that currents only occur at the edges, which since only
here $E_{n,Y}$ is an explicit function of $Y$, see Fig.\ 4. 

We now open the system by cutting 
the periodic boundaries in the $x$ direction and connect the
Hall bar to two reservoirs. One is kept at a chemical potential $\mu_A$,
and the other at a chemical potential $\mu_B$.  This is shown in figure
Fig.\ 5.  What is the total current flowing from one reservoir to the 
other one through the Hall bar?  We start by only considering the current
associated with the $n$th Landau level, $I_{x,n}$.  This current is 
$$
I_{n,x}=\int_{\mu_A}^{\mu_B} d\mu\ e v_x(n,Y(\mu)) g_n(\mu)\;,
\eqno(65)
$$
where we use the one-dimensional density of states propagating to the 
right,
$$
g_n(E)={1\over L}\ {{dj_n(E)}\over{dE}}
= {{1}\over{2\pi l^2}}\ {{dY_n(E)}\over{dE}}\;.
\eqno(66)
$$
The reason for using the {\it one-dimensional\/} expression, is that
the edge currents are one-dimensional.  We have used Eq.\ (22) in going 
from $j$ to $Y$ on the right hand side of Eq.\ (66).  The limits 
of the integral in Eq.\ (65) reflect that it is the excess current we are
measuring, as the currents associated with matching chemical potentials 
cancel (but they flow along opposite edges).  Combining these
two last equations with Eq.\ (64), we find that the total current in
the $n$th Landau level is
$$
I_{n,x}={e\over h}\left(\mu_B-\mu_A\right)\;.
\eqno(67)
$$
These currents are, as we have seen, edge currents.  There is one such 
current for every bulk Landau level which is below the Fermi level.  The
reason for this may be seen in Fig.\ 4:  The Landau levels bend
upwards at the edges, and those below the Fermi level in the bulk cut through
the Fermi level at the edges.  Thus, the total current is, when both $\mu_A$ 
and $\mu_B$ differ infinitesimally from the Fermi level,
$$
I_x=N\ {e\over h}\left(\mu_B-\mu_A\right)\;,
\eqno(68)
$$
where $N$ is the number of bulk Landau levels below the Fermi level.
It is an integer. 

The factor $Ne/h$ in Eq.\ (68) is an inverse {\it contact resistance\/}
(divided by $e$), meaning that it is due to a redistribution of the 
population 
of states when the charges enter the system from that they had in the 
reservoirs. If one measures the voltage difference between contacts 1 and 2, 
or 3 and 4, which are shown in Fig.\ 5, one finds zero: There is
no change in chemical potential along the edges.  However, measuring 
voltage difference across the system, i.e., between contacts 1 and 3,
1 and 4, or 2 and 3, or 2 and 4, one finds a 
potential difference $\mu_B-\mu_A$.
Thus, there is no resistance in the flow direction of the current, and there
is a resistance in the direction perpendicular to the flow direction.  In
other words, we have
$$
\rho_{xx}=\rho_{yy}=0\;,
\eqno(69)
$$
and\footnote{The extra $1/e$ factor in Eq.\ (70) comes from
measuring chemical potential --- which has the units of energy --- in
units of electric potential.}
$$
\rho_{yx}=-\rho_{xy}={h\over{e^2 N}}\;.
\eqno(70)
$$
\begin{figure}[ht]
\vspace*{7.3cm}
\includegraphics{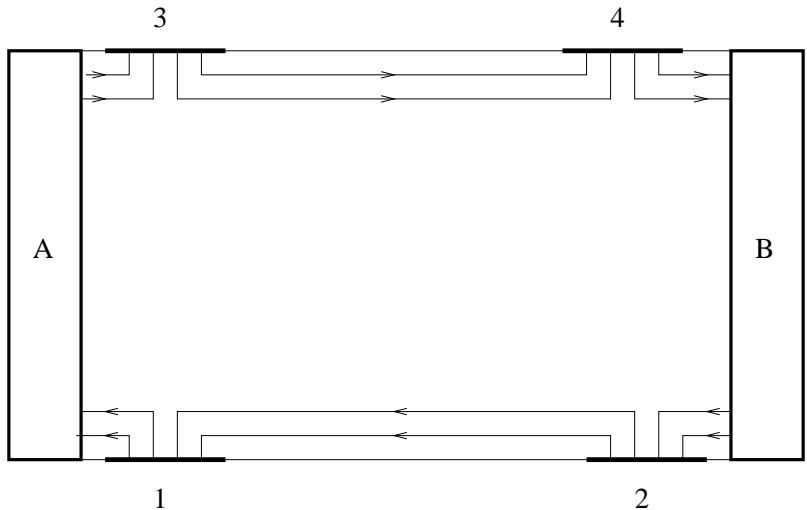}
\caption{A clean Hall bar (i.e., $V=0$) in contact 
with two reservoirs $A$ and $B$ kept at chemical potentials $\mu_A$ and 
$\mu_B$.  Furthermore, four contacts have been added so that the 
potential between any pair among them may be measured.  There is no current 
passing through contacts 1, 2, 3, and 4.}
\end{figure}
These resistances should be compared to those listed in Eqs.\ (43) and
(44).  The difference between the two sets of resistances occurs
in Eqs.\ (70) and (44):  In Eq.\ (70), the Hall resistance is
quantized, while in Eq.\ (44), it is not.  

Why this difference?  In Sec.\ 2, we worked with a closed system.
The number of electrons was fixed.  The present system is open; it is in
contact with two reservoirs.  Whenever a new Landau level drops below the
Fermi level, e.g.\ through decrease of the magnetic field $B$, 
it fills up due to the contact with the reservoirs.  This is not possible
in the closed system. 

We now ``turn" on the disordered potential $V$. The arguments leading to the 
bending upwards of the Landau levels at the edges of the clean Hall bar
rested on there being sharp boundaries on the scale of the magnetic length
$l$, cutting the wave function off.  It is more realistic to assume that
the boundaries are sharp on the scale of the fluctuations of the 
disorder potential $V$, which in strong magnetic fields is much larger
than the magnetic length $l$.  Thus, at the edges, the potential $V$ turns
smoothly upwards on the scale of $l$, 
while the Landau levels remain unchanged throughout the
system.\footnote{The conclusions we draw in the following are 
valid even when the edges are sharp on the scale of $l$ --- see 
Halperin [25].}

We have seen in Sec.\ 3 that the electrons follow equipotential
curves in the energy landscape.  In the bulk of the Hall bar, most such
curves form closed loops circling ``mountains" or ``valleys" in the 
energy landscape.  However, at the edges, there are equipotential paths
reaching from one end of the sample to the other due to the upwards bend
of the energy eigenvalues.  Connecting the Hall bar with reservoirs $A$
and $B$ kept at chemical potentials $\mu_A$ and $\mu_B$ will induce currents
along these paths.  How many such paths are there?  The argument is exactly  
the same as in the case with $V=0$ (i.e., the clean case):  There
is one path for every Landau level that is below the Fermi level in bulk.

It is useful to picture the following situation:  Take a plate
and with a hammer knock it into a shape that resembles our disorder potential.
At the edges, bend the plate upwards.  This correponds to the bending
at the edges of the Landau levels.  Then make a series of copies of 
this plate and stack them at regular height intervals.  This is a 
three-dimensional model of the energy levels in the Hall bar.  Now, place the
whole construction in a tub and start filling it with water.  The water
level represents the Fermi level.  As the water rises, ``shore lines" will 
form where the dry plates dives into the water. These shore lines correspond 
to the equipotential curves at the Fermi level.  We see that for most water 
levels, shore lines run parallel to upwardly bent parts of the plates but
not across the plates, orthogonal to the bends.  However, for certain levels,
there {\it are\/} connected shore lines running across the sample in 
connecting the two bent edges.  This happens every time a new plate 
(i.e., Landau level) sinks into the water (i.e., the Fermi 
sea). 

Returning from the tub to the Hall bar, how does this appearance of paths
running across the sample parallel to the $y$ direction (i.e., from
edge to edge) affect the conductances of the system?  Before we
answer this question, we must determine the conductances of the system
as long as such paths are {\it not\/} present.  In order to argue this,
we note that the arguments leading to Eq.\  (68) are more general
than the derivation seems to indicate.  The point is that this conductance
is, as already pointed out, a {\it contact\/} conductance caused by 
the redistribution of the Fermi distribution in the reservoirs at the 
contacts between the reservoirs and the conductor.  B{\"u}ttiker [24] 
therefore 
simply assumes that each current channel (i.e.,  edge state) is a 
one-dimensional perfect conductor.  The Fermi energy is in the conductor 
$E_F=E_n+\hbar^2k^2/2m$, where $m$ is the effective mass of the electrons.
The velocity of the electrons is $v_n=dE_n/dk$.  The density of states
$g_n(E)=d{\cal N}_n/dE=(1/2\pi)dk/dE$, since $d{\cal N}_n/dk=1/2\pi$ 
in one dimension.  Thus, the combination $ev_n(E) g_n(E)=
e(dE/dk)(dk/dE)(1/2\pi\hbar)=e/h$.  The current in channel $n$ is
$I_n=\int_{\mu_A}^{\mu_B} d\mu ev_n(\mu)g_n(\mu)=(e/h)(\mu_B-\mu_A)$,
i.e., Eq.\  (68).  As a result, we find longitudinal and
transversal (i.e., Hall) resistances to be those listed in Eqs.\
(69) and (70).\footnote{B{\"u}ttiker [24] 
discusses {\it inelastic\/} 
scattering, and shows that even in this case, the resistivities will be given 
by Eqs.\ (69) and (70) as long as there are no scattering paths 
connecting the edges across the sample.}

In Fig.\ 6, we show a ``dirty" Hall bar connected to two reservoirs
$A$ and $B$, and four contacts 1, 2, 3, and 4.  Following the 
Landauer-B{\"u}ttiker formalism for phase coherent transport [26],
we write down the balance equations for the 6-terminal system of Fig.\
6.  There is a current $I_i$ associated with contact $i$ in this 
system.  The sign is chosen so that it is positive if it enters the Hall bar.  
We measure the chemical potentials at the reservoirs $A$ and $B$ in units
of voltage, $\mu_A=eV_A$ and $\mu_B=eV_B$.  We assume that the Fermi level
is set so that $N$ Landau levels are beneath it, while Landau level number
$N+1$ has equipotential curves cutting across it from edge to edge at the
Fermi energy.  There is a transmission coefficient $T$ associated 
with reaching contact 1 from contact 2 and reaching contact 4 from contact 3,
and a reflection coefficient $R$ associated with reaching contact 4 from
contact 2 and with reaching contact 1 from contact 3.  Furthermore, $R+T=1$.
Due to the symmetry
\begin{figure}[t]
\vspace*{7.3cm}
\includegraphics{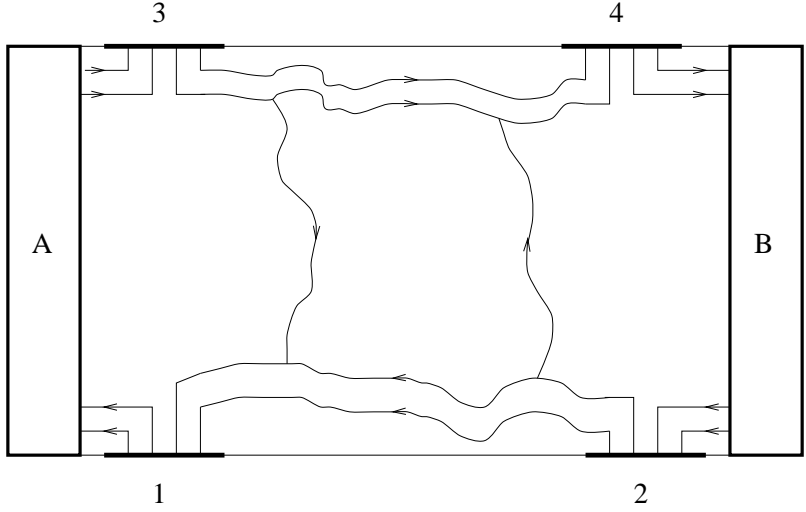}
\caption{A dirty Hall bar (i.e., $V\neq 0$) in contact 
with two reservoirs $A$ and $B$ 
kept at chemical potentials $\mu_A$ and $\mu_B$.  Four contacts
have been added as in Fig.\ 5.}
\end{figure}
of the problem under a change of direction of the 
magnetic field, $\vec B\to-\vec B$, there is only one reflection and 
transmission coefficient, and not two as in the more general case.
We write down the balance equations,
\setcounter{equation}{70}
\begin{eqnarray}
I_A & = & {{e^2}\over h}\left((N+1)V_A-(N+1)V_1\right)\;,
\\
I_B & = & {{e^2}\over h}\left((N+1)V_B-(N+1)V_4\right)\;,
\\
I_1 & = & {{e^2}\over h}\left((N+1)V_1-(N+T)V_2-RV_3\right)\;,
\\
I_2 & = & {{e^2}\over h}\left((N+1)V_2-(N+1)V_B\right)\;,
\\
I_3 & = & {{e^2}\over h}\left((N+1)V_3-(N+1)V_A\right)\;,
\end{eqnarray}
and
\begin{equation}
I_4={{e^2}\over h}\left((N+1)V_4-(N+T)V_3-RV_2\right)\;.
\end{equation}
The experimental configuration that allows one to measure the longitudinal
and Hall resistances is one where
$$
I_1=I_2=I_3=I_4=0\;.
\eqno(77)
$$
It immediately follows that
$$
I_A=-I_B=I\;.
\eqno(78)
$$
We now choose one of the six potentials equal to zero, e.g.\ 
$V_B=0$, to anchor the voltage scale.  Equations (71) -- (76) 
may then be transformed into
\setcounter{equation}{78}
\begin{eqnarray}
I & = & {{e^2}\over h}(N+T)V_A\;,
\\
V_1 & = & {{e^2}\over h}{{R}\over{N+1}}V_A\;,
\\
V_2 & = & 0\;,
\\
V_3 & = & {{e^2}\over h}V_A\;,
\end{eqnarray}
and
\begin{equation}
V_4={{e^2}\over h}{{N+T}\over{N+1}}V_A\;.
\end{equation}
We may now determine the resistances between all pairs of contacts:
\begin{eqnarray}
R_{34}={{V_3-V_4}\over{I}} & = & {h\over{e^2}}{{R}\over{(N+1)(N+T)}}\;,
\\
R_{12}={{V_1-V_2}\over{I}} & = & {h\over{e^2}}{{R}\over{(N+1)(N+T)}}\;,
\\
R_{42}={{V_4-V_2}\over{I}} & = & {h\over{e^2}}{{1}\over{(N+1)}}\;,
\\
R_{31}={{V_3-V_1}\over{I}} & = & {h\over{e^2}}{{1}\over{(N+1)}}\;,
\\
R_{41}={{V_4-V_1}\over{I}} & = & {h\over{e^2}}{{N-1-2T}\over{(N+1)(N+T)}}\;,
\end{eqnarray}
and
$$
R_{32}={{V_3-V_2}\over I}={h\over{e^2}}{{1}\over{(N+1)}}\;.
\eqno(89)
$$
If now $T=1$ and $R=0$, which is the case when there is no path across the
sample at the Fermi level, Eqs.\ (84) to (89) reduce to
$$
R_{12}=R_{34}=0\;,
\eqno(90)
$$
and
$$
R_{42}=R_{31}=R_{41}=R_{32}={h\over{e^2}}{1\over{N+1}}\;.
\eqno(91)
$$
This is the integer quantum Hall effect, and we recognize Eqs.\ 
(69) and (70) in the last two equations.

From Eqs.\ (84) to (89) we also see what happens when
the system moves from one Hall plateau, i.e., when Eqs.\ (90)
and (91) reign, to the next:  A longitudinal resistance appears,
as long as $R=1-T>0$.  This is the localization-delocalization transition
that appears in the Introduction to 
this article.  And, we see now what causes it,
namely the appearance of paths across the sample in the $y$ direction
at the Fermi level.  But, what kind of transition is this?  In the early
eighties it was suggested that it is the same mechanism as that of 
coalescing ponds during rainy weather, i.e., it is a 
{\it percolation transition,\/} see Kazarinov and Luryi [27] and 
Trugman [28].  However, as has now become clear, this picture was too
simple minded.  

\bigskip
\centerline{\bf 4.\ The Critical Points}
\bigskip
After having set the scene, we are now ready to discuss the phenomenon
which is the aspect of the integer quantum Hall effect that forms the 
focus of this review:  The critical behavior associated with changing
the quantized filling factor from $N$ to $N\pm 1$. 

\bigskip
\centerline{\bf 4.1 Expectations and Observations}
\bigskip
As stated at the end of Sec.\ 3, it was early theorized that the 
critical behavior is that of ``standard" percolation.  The theory goes
as follows:

There are three classes of electron wave functions in the system: (i) Wave 
functions localized to closed paths that follow the sides of ``valleys'' in 
the energy landscape $V$, (ii) those localized to closed paths circling the 
``mountains'', and (iii) those localized to the group of paths that consists 
of paths of type (i) or (ii) and which are connected through saddle points.  
For one particular level $V({\vec x})=constant$, there is at least one 
connected 
cluster of paths (i.e., a separatrix of type (iii)) that spans 
the system, connecting opposite edge states.  The corresponding energy level 
we denote $E_{c,n}$.  As explained in Sec.\ 3.2, a longitudinal  
conductance appears each time the Fermi level passes through an $E_{c,n}$
by for example adjusting the perpendicular magnetic field $B$.  There is
a particular magnetic field $B_{c,n}$ corresponding to the energies
$E_{c,n}$.

The typical (linear) size of area the connected wave functions span, $\xi_p$, 
diverges as\footnote{The last proportionality in this equation is
true as long as $E=E(B)$ is an analytic function near $B_{c,n}$, defined
as $E_{c,n}=E(B_{c,n})$.} 
$$
\xi_p\propto|E-E_{c,n}|^{-\nu}\propto |B-B_{c,n}|^{-\nu}\;,
\eqno(92)
$$
where the percolation connectivity length exponent $\nu$ is known to be
$4/3$ in two dimensions.  
The localization length, $\xi$, is the length scale over which the   
wave function does not fall off faster than algebraically, and behaves as
$$
\xi\propto|E-E_{c,n}|^{-\tilde\nu}\;,
\eqno(93)
$$
where $\tilde\nu$ 
is the localization length exponent.  If percolation, in the 
sense described above, were the only mechanism, we would have
$$
\tilde\nu=\nu={4\over 3}\;,\quad {\rm wrong!}
\eqno(94)
$$
This, however, is {\it not\/} what is observed in experiments, nor 
in computer simulations. Koch {\it et al.\/} [29,30] measured 
$\tilde\nu=2.3\pm 0.1$, while Wei {\it al.\/} [31,32] found the 
value $\tilde \nu=2.4\pm0.2$. This value was not measured directly, but 
relies on assumed values of other exponents.  On the other hand,
Shashkin {\it et al.\/} [33,34] and Dolgopolov {\it et al.\/} [35] observe
$\tilde\nu\approx 1$, which is not too far from                  
the percolation value $\nu=4/3$. However, also in these three latter papers, 
$\tilde\nu$ was not measured directly, but relied on theoretical arguments. 

Several numerical simulations have been performed to determine $\tilde\nu$.  
Models based on continuum descriptions of the problem, using long-wavelength 
Gaussian potentials or short-range impurity potentials, result in 
$\tilde\nu\approx 2.3$ -- 2.4 [36--41]. 

Network models, originally introduced by Chalker and Coddington [42] give a 
similar value for $\tilde\nu$, see [43,44]. We describe these network models
in detail in Sec.\ 5.

These for the most part mutually consistent results strongly hint at the
localization-delocalization transition not being describable by 
classical percolation alone.  Quantum effects play a r{\^o}le.  

There are two ways that quantum mechanics may change 
the universality class of the transition, while still keeping the physical
picture developed in the preceeding sections of this review and leading to
the percolation picture just presented: 1) Tunneling at the saddle points
and 2) interference at the saddle points, making this an example of 
Anderson localization.

\bigskip
\centerline{\bf 4.2 The Mil'nikov and Sokolov Argument}
\bigskip
There is a famous argument, first presented by Mil'nikov and Sokolov [20], 
that leads to 
$$
\tilde\nu=\nu+1={7\over 3}\;,
\eqno(95)
$$
based on purely semi-classical reasoning, i.e., taking only tunneling into 
account while ignoring interference.  

We follow first Hansen and L{\"u}tken [45] in their interpretation of the 
Mil'nikov--Sokolov argument.  A different approach, which we also sketch, 
can be found in Zhao and Feng [46].

Let us assume that there is a typical distance $a$ between saddle points.  
We tune the magnetic field, and find classical overlap at saddle points in a 
window $\Delta B$ around $B_{c,n}$ which creates a connected cluster across 
the sample. Tunneling changes the question of overlap: Overlap is then no 
longer characterized as overlap within the magnetic length $l$, but by the
tunneling probability $T$ given by Eqs.\ (58) and (59).  Let us write Eq.\ 
(59) as
$$
P=e^{-1/\chi}\;,
\eqno(96)
$$
where
$$
\chi={{ml^2\sqrt{\alpha\beta}}\over{he |\Delta B|}}\;.
\eqno(97)
$$
Classically, we may define an ``elementary" localization length as    
the typical distance between saddle points, $a$.  This is the length over 
which the electrons may move if there are no overlaps at the saddle points,
as they are constrained to follow closed paths with this average diameter.

When tunneling is included, the elementary localization length is changed.
When $\Delta B$ is small, we have from Eqs.\ (58) and 
(97) that $T\approx e^{-1/\chi}$.  An electron starting somewhere in the 
system will pass $K$ such saddle points with a probability $e^{-K/\chi}$ 
(when interference effects are ignored).  Fixing this probability to, say, 
$e^{-1}$, we may define a {\it tunnel length\/} to be 
$$
\Lambda= Ka  = \chi a \;.
\eqno(98)
$$
From Eq.\ (97) we see that $\Lambda \sim |\Delta B|^{-1}$ when 
$B\to B_c$. In terms of the percolation picture, $\Lambda$ takes over from
$a$ as the lower cutoff in length scale when tunneling is taken into account,
an assumption relying on the one-dimensional character of the percolation
clusters near threshold.\footnote{The percolating cluster has the 
topology of a one-dimensional chain with ``blobs" interspersed among the 
so-called ``cutting bonds."  The cutting bonds are those that cuts off 
percolation if removed.  The blobs  are clusters of connected bonds.  
Coniglio [47] has shown that the fractal dimension of the cutting bonds is 
$1/\nu$ at the percolation threshold.} The percolation correlation length 
$\xi$ is essentially the size of the largest cluster of connected ribbons. 
Measured in units of the effective elementary localization length, 
$\Lambda=\chi a$, $\xi$ diverges as 
$$
{{\xi}\over{\Lambda}}={{\xi}\over{\chi a}}\sim |\Delta B|^{-\nu}\;, 
\eqno(99)
$$
as relative to this length, the problem is that of ``ordinary" percolation.  
However, measured relative to a fixed length, such as $a$, one has that 
$$
{{\xi}\over{a}}\sim |\Delta B|^{-(\nu+1)}=|\Delta B|^{-\tilde\nu}\;,
\eqno(100)
$$
and Eq.\ (95) follows.

Equation (99) implies that tunneling {\it increases\/} the localization
length in comparison to classical percolation, i.e.,
$$
\xi>\xi_p\;.
\eqno(101)
$$
However, Zhao and Feng [46] argues that 
$$
\xi\ll\xi_p\;.
\eqno(102)
$$
Their argument goes as follows: We are close to, but not at the classical
percolation threshold, and have no infinite cluster yet.  
The largest cluster of connected paths that would 
emerge at the percolation threshold has therefore been broken up into
islands of size $\xi_p$.  Between each island there is a saddle point where
tunneling is necessary.  The system has a width $W$, and therefore 
the number of such tunneling saddle points is $W/\xi_p$.  The probability
that an electron tunnels through all these saddle points along a
path crossing the sample is therefore
$$
P_W=P^{W/\xi_p}=e^{-W/(\xi_p\chi)}=e^{-W/\xi}\;,
\eqno(103)
$$
where we have used Eq.\ (96).  The right hand expression defines the 
localization length, and we find
$$
\xi=\chi\xi_p\sim |\Delta B|^{-(\nu+1)}\;.
\eqno(104)
$$
We have here used Eqs.\ (92) and (97).  From Eq.\ (97), we see that
a tunnel length scale may be defined,
$$
l_t=\sqrt{{he|\Delta B|}\over{m\sqrt{\alpha\beta}}}\;,
\eqno(105)
$$
and Eq.\ (104) may be written
$$
\xi=\xi_p\left({l\over{l_t}}\right)^2\;.
\eqno(106)
$$
Since tunneling has been assumed to be the dominating mechanism for transfer
across the saddle point, one must have
$$
l_t\gg l\;.
\eqno(107)
$$
Combination of this 
inequality with Eq.\ (106) leads to the inequality (102).  It should be noted
here that as $|\Delta B|\to 0$, this argument implies a crossover to 
classical percolation behavior when $l_t$ becomes of the order of $l$. More
detailed analytical arguments [48] have been constructed that support the
rescaling argument in one dimension, but {\it not\/} in two dimensions, which
is our concern here.  On the other hand, no crossover as implied by the 
argument basic to Eq.\ (106) is found in numerical work.
                                
In the above arguments, quantum interference has been left out.  It would
lead to an Anderson-type localization [1].  There are no arguments on the
level of the Mil'nikov-Sokolov that predicts a value for the localization
length exponent $\tilde\nu$ based on interference.  In order to investigate the
implications of this mechanism, numerical studies are required.  We therefore
now turn to the numerical modelling of the integer quantum Hall problem.

\bigskip
\centerline{\bf 5.\ Numerical Models}
\bigskip
In this Section, we turn to the numerical studies of the integer quantum 
Hall effect.  Since Mil'nikov-Sokolov-type argumentation is controversial,
and no arguments taking the full quantum mechanical formulation into account 
that are capable of pinning down the value of $\tilde\nu$, 
it seems that the computer 
will continue to be the principal tool for studying this problem.  The use 
of numerical methods has, indeed, been quite successful in this field.  Not 
only have results found in experiments been well reproduced on the computer,
but also computer results have had predictive power. For example, the 
Chalker-Coddington model [42], which we will be studying in a moment,  
predicted in 1988 the value\footnote{Reanalysing the original
data of Chalker and Coddington [42] using a different parame\-trization 
and including additional data, Huckestein [39] found 
$\tilde\nu=2.35\pm0.03$.} $\tilde\nu=2.5\pm 0.5$, while the experimental 
result $\tilde\nu=2.3\pm0.1$ of Koch {\it et al.\/} [29], was reported
three years later.

\bigskip
\centerline{\bf 5.1\ The Chalker-Coddington Model}
\bigskip
We will in this review concentrate on network models based on the one
introduced by Chalker and Coddington [42], as it seems that these models
capture very well the critical
\begin{figure}[t]
\vspace*{8.2cm}
\includegraphics{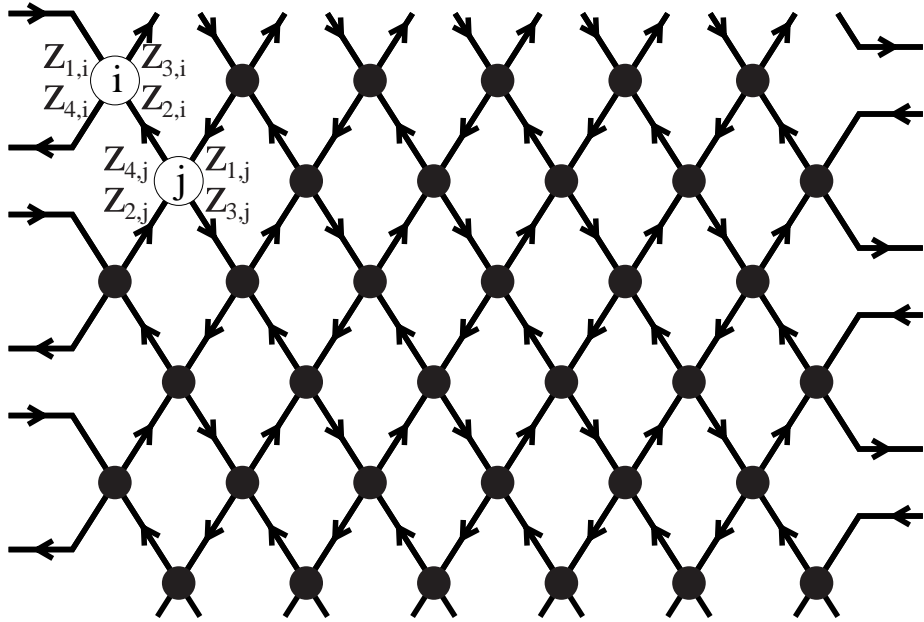}
\caption{Network used in the Chalker-Coddington model.  It
has length $L$ in the horizontal direction and length $M$ in the vertical 
direction.  It is periodic in the vertical direction.  The amplitudes
associated with nodes $i$ and $j$ are shown.}
\end{figure}
aspects of the integer quantum Hall effect.  
Other models may be found in Huckestein's review [13].

The Chalker-Coddington model replaces the random network of paths in
the disordered energy landscape by a regular square
lattice with a lattice constant
equal to the average distance between saddle points in the disordered
energy landscape.  The nodes of the lattice represent saddle points, and the 
bonds the paths between the saddle points.  Each bond can only be traversed 
in one direction.  The possible directions are chosen so that the network 
forms a grid of loops of alternating handedness, see Fig.\ 7. The wave 
function itself is represented by a complex variable $z$.  At a given saddle 
point, $z_1$ and $z_2$ are ingoing amplitudes, and $z_3$ and $z_4$ outgoing 
amplitudes as shown in Fig.\ 8. The scattering matrix is conventionally 
written\footnote{Traditionally, the scattering matrix is written 
in the form (108).  It could equally well have been written in a form that
relates the outgoing channels $(z_3,z_4)$ to the ingoing channels 
$(z_1,z_2)$.} [42]
$$
\left( \matrix {z_1\cr z_3\cr}\right)=
\left(\matrix {e^{i\phi_1} & 0 \cr 0 & e^{i\phi_2} \cr} \right)
\
\left(\matrix{\cosh\gamma & \sinh\gamma \cr 
\sinh\gamma & \cosh\gamma \cr}\right)
\ 
\left(\matrix{e^{i\phi_3} & 0 \cr 0 & e^{i\phi_4} \cr}\right)
\
\left(\matrix{ z_4\cr z_2\cr} \right)\;. 
\eqno(108)
$$
where $\phi_1$, $\phi_2$, $\phi_3$ and $\phi_4$, represent the phase changes 
of the wave function along the bonds. These are chosen at random.  This
randomness represents the variation in path length from loop to loop.  The
control parameter $\gamma$ determines the scattering properties at the
saddle point. When $\gamma\to 0$,
\begin{figure}[t]
\vspace*{7cm}
\includegraphics{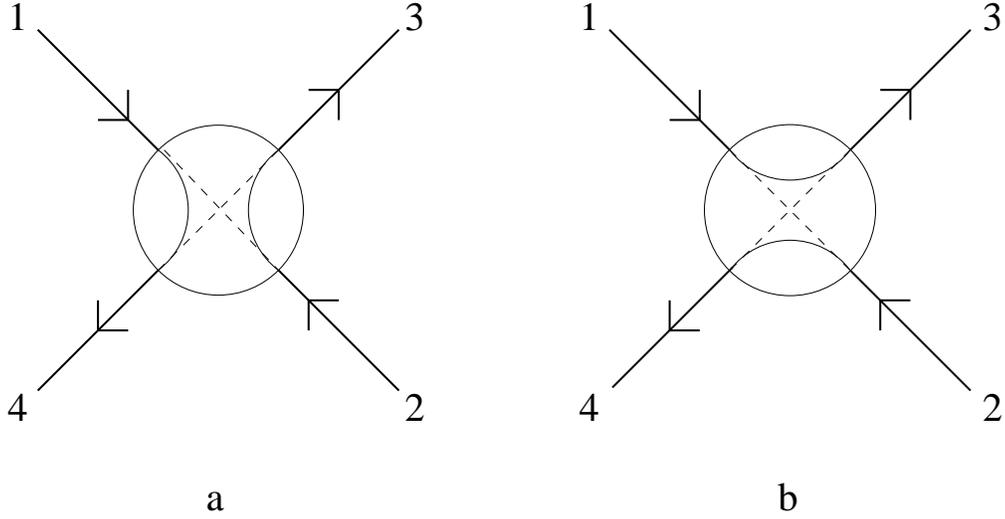}
\caption{Scattering at a given node in the Chalker-Coddington model: 
a) shows the case when $\gamma\le\gamma_c$ while b) shows the case 
$\gamma\ge\gamma_c$.  The dashed curves denote the case $\gamma=\gamma_c$.}
\end{figure}
the transmission probabilities are
$P_{1\to 4}=P_{2\to 3}=1$ and $P_{1\to 3}=P_{2\to 4}=0$.  Likewise,
when $\gamma\to\infty$, we have that $P_{1\to 3}=P_{2\to 4}=1$ and 
$P_{1\to 4}=P_{2\to 3}=0$.  When $\gamma=\gamma_c=\ln (1+\sqrt{2})$, one
finds $P_{1\to 4}=P_{1\to 3}=P_{2\to 4}=P_{2\to 3}=1/2$.  When 
$\gamma<\gamma_c$, $P_{1\to 3}=P_{2\to 4}=1/(1+\sinh^{-2}\gamma)$ are
due to tunneling, while $P_{1\to 4}=P_{2\to 3}=1-P_{1\to 3}$ are direct
channels.  When $\gamma>\gamma_c$, $P_{1\to 4}=P_{2\to 3}=
1/(1+\sinh^{+2}\gamma)$ are due to tunneling 
while the two remaining channels
are direct.  This is illustrated in Fig.\ 8.  Writing $P$ as
in Eq.\ (96), derived in the energy landscape model, we may 
combine all these results in the expression
$$
\chi={1\over{2|\ln\sinh \gamma |}}\;.
\eqno(109)
$$

While Chalker and Coddington [42] kept $\gamma$ equal for all nodes, with 
disorder in the phases $\phi_i$ only, Lee {\it et al.\/} [43] extended the 
original model by writing 
$$
\gamma=\gamma_c e^{\mu-v}\;, 
\eqno(110)
$$
thereby introducing disorder also in the saddle points in addition to the one 
already present.  Here $\mu$ is a fixed parameter, equal for all nodes in the 
network, while $v$ is chosen randomly on the interval $[-w/2,+w/2]$, where
$w$ is a parameter determining the width of the distribution. As we will  
see in Sec.\ 5.2, the Chalker-Coddington model crosses over to classical
behavior in the limit $w\to\infty$. In the 
case when $|\mu-v|\ll 1$, we may relate $|\mu-v|\propto |B-B_c|$ of the 
energy landscape model.  By using Eq.\ (98) combined with 
Eqs.\ (109) and (110), we find that the tunnel 
length $\Lambda$ behaves as 
$$
\Lambda= \chi a\propto {a\over{|\mu-v|}}\;,
\eqno(111)
$$
with $a$ the lattice constant.  This expression should be compared to the
corresponding expression, Eq.\ (97), derived in the smooth-potential model.
Thus, the renormalization argument presented in Sec.\ 4.2, predicts the 
same critical behavior in the Chalker-Coddinton model as in the energy 
landscape model.  However, we reiterate that a localization length exponent
$\tilde\nu=2.35\pm 0.03$ [39] is found in the Chalker-Coddington model 
{\it without\/} nodal disorder.  This has been investigated further,
with the conclusion that the Chalker-Coddington model does 
{\it not\/} support the Mil'nikov and Sokolov argument [49].

The way one determines $\tilde\nu$ in the Chalker-Coddington model is based 
on a transfer matrix approach and the assumtion of one-parameter scaling 
[50].  The procedure is as follows.  The lattice employed is as shown in
Fig.\ 7, but with horizontal length $N$ and vertical length $M$.  We assume
$N\gg M$ --- typically $M$ is of the order of $10^2$ and $N$ of the order
$10^5$. Calculate the trace of the transmission probability from layer one
to layer $N$, ${\rm tr}|T|^2$. Average this quantity {\it geometrically\/}  
over the different samples, i.e., calculate 
$\exp[\langle \ln {\rm tr}|T|^2 \rangle]$.  The reason for doing the 
geometric average is that we want the typical transmission probability.  This
is not accessible with the usual arithmetic average, 
$\langle {\rm tr}|T|^2 \rangle$, which is dominated by the rare events where
the sample actually percolates in the classical sense from edge to edge (in
the $M$ direction), and in which case ${\rm tr} |T|^2$ is equal to one.
In taking the logarithm, these rare events no longer play a r{\^o}le.
We may now define an effective quantum localization length, $\xi_M$ by
$$
{1\over{\xi_M}}=-\lim_{N\to\infty} {1\over N}\ 
\langle \ln {\rm tr}|T|^2\rangle\;.
\eqno(112)
$$
It is effective in the sense that it depends on $M$.  The asymptotic
localization length, which is the one we want to determine, is then
$$
\xi=\lim_{M\to\infty}\xi_M\;.
\eqno(113)
$$
At this point, the one-parameter scaling assumption comes into play.  It
assumes the following functional relationship between $\xi$ and $\xi_M$
with $M$ and $\mu$, defined in Eq.\ (110), as parameters: 
$$
{{\xi_M}\over M}=F\left({{\xi(\mu)}\over M}\right)\;.
\eqno(114)
$$
By plotting $\xi_M$ for several values of $\Delta B$ and
$M$, data collapse is obtained when the correct functional dependence
of $\xi$ on $\Delta B$ has been found --- which is a power law with slope
$-\tilde\nu$.

We shall not discuss the basis for the above procedure here, but instead
turn to a different way of extracting the localization length exponent from
the Chalker-Coddington model.  One recently employed consists in treating 
the network as an {\it open\/} system [51].  By open, we mean the 
following:  Rather than calculating transmission and reflection 
coefficients from the elementary scattering matrices of Eq.\ (108),
we ``hook" the network up to a source and a sink.  That is, we specify the
value of $z$ on all bonds on the edges leading into the network, while 
leaving the $z$ values of the exiting bond unspecified.  This corresponds to 
driving a current through the network, and $z$ must now be interpreted as 
current amplitudes rather than wave functions, as it is no longer normalized. 
In fact, this models the dirty Hall bar with reservoirs shown in Fig.\ 6, 
although the boundary conditions in the transverse direction are different.
$\tilde\nu$ can then be determined as is done in connection with the 
classical random resistor network at the percolation 
threshold.\footnote{Suppose the bonds of a standard bond percolation
problem are electrical resistors, all equal.  What are the electrical
properties of this network?  First of all, all the geometrical and 
topological properties of networks at the percolation threshold are still
in place; this is a transport problem on top of the original percolation
problem.  In particular $\nu=4/3$ in two dimensions.  In addition, one
finds that global quantities such as the conductance $G$ scale with system
size $L$ as 
$$
G\sim L^{-t/\nu}\;,
$$
where $t=1.300$ [52].  
Its exact value is not known.  Furthermore, the current
distribution is {\it multifractal.\/}  We will explain this concept below. 
See Ref.\ [53] for a review.}

When the system is considered open, the various link amplitudes are found by
solving a set of linear equations,
$$
Az=b\;,
\eqno(115)
$$
where $B$ contains the $z$ values that correspond to specified bonds, and the
matrix $A$ contains the detailed scattering conditions for each node, based
on Eq.\ (108).  We fix all currents entering the network shown in Fig.\
7 from the left to unity, while no currents enter the network from the right.
By solving Eq.\ (115), we determine all $z$-values in the network, and from
these the total current passing through the system, namely,
$$
I=\sum_{i\ \in\ {\rm first\ column}} 
\left(|z_{1,i}|^2-|z_{4,i}|^2\right)=\sum_{i\ \in\ {\rm last\ column}}
|z_{3,i}|^2\;.
\eqno(116)
$$
The first of the two lower indices on the bond variables $z$ refers to the 
scattering channel as shown in Fig.\ 7, while the second index is the 
node address.  

The total current entering the network from the left, $I_l$ is 
proportional to the chemical potential of the reservoir hooked up to the 
network on that side, while the chemical potential on the right is 
zero, as no current enters from this side.  Thus, the potential difference 
across the network is simply $V=h/e^2 I_l$. We may now define a longitudinal
conductance $G$ through the expression
$$
I=V G\;.
\eqno(117)
$$

As is evident from Fig.\ 7, the nodes belonging to every second row have 
their scattering labels rotated by $90^\circ$ in the clockwise direction.
As a result of this rotation, the connection between bond labels and
forward/backward scattering depends on which row the nodes belong to.
For the odd-numbered rows, scattering $z_1\to z_3$ represents forward
scattering, whereas for the even rows, it is represented by $z_2\to z_3$. 
The scattering amplitude for channel $z_1\to z_3$ 
increases with increasing $\gamma$, whereas the scattering amplitude for
channel $z_2\to z_3$ increases with decreasing $\gamma$.        

Rather than assigning random values for $\gamma$ to the nodes in the network,
we have used two bimodal distributions, one for the odd rows,
$$
p_o(\gamma)=(1/2+\mu)\delta(\gamma-\gamma_{13})+
(1/2-\mu)\delta(\gamma-\gamma_{23})\;,
\eqno(118)
$$
and one for the even rows,  
$$
p_e(\gamma)=(1/2-\mu)\delta(\gamma-\gamma_{13})+
(1/2+\mu)\delta(\gamma-\gamma_{23})\;.
\eqno(119)
$$
These two distributions contain three parameters: $\gamma_{13}$, 
$\gamma_{23}$ and $\mu$.  We relate $\gamma_{13}$ and $\gamma_{23}$  
through the relation
$$
\tanh \gamma_{13} = {1\over{\cosh \gamma_{23}}} =t_0\;,
\eqno(120)
$$
where $t_0$ is a transmission amplitude. With the relation (120),
a node belonging to an even row with a $\gamma=\gamma_{23}$ will
transmit in exactly the same manner as a node belonging to an odd
row with $\gamma=\gamma_{13}$.  When $|t_0|\to 1$, 
tunneling becomes less and less likely and the model becomes classical,
as will be discussed in Sec.\ 5.2.  On the other hand, in the 
$|t_0|\to 1/\sqrt{2}$ limit we find that $\gamma_{13}=\gamma_{23}=\gamma_c
=\ln(1+\sqrt{2})$.

With a large negative value of control parameter $\mu$, the system will
predominantly consist of elementary loops with current running clockwise,
whereas with a large positive value of $\mu$, the current will be running 
counterclockwise in loops shifted by one lattice constant in the (11) 
direction.  When $\mu=0$, there is no bias and the system is critical.

In our simulations, we set $t_0=0.9$, and let $\mu$ vary between -0.5 and
0.5.  In Fig.\ 9 we show the current $I$ as a function of $\mu$ for square 
system sizes $L=12$ to 92.  The data were geometrically averaged over 14650 
samples for $L=12$ to 300 samples for $L=92$.

From Fig.\ 9 we see that the larger the system size, the sharper the
peak. The system becomes critical when $L/\xi(\Delta \mu) \simeq 1$, hence
we find that the half width, $\Delta \mu$, scales as
$$
\Delta \mu \propto L^{-1/\tilde\nu}\;.
\eqno(121)
$$ 
By plotting the half width $\Delta \mu$ versus system size in a log-log
plot, Fig.\ 10, we obtain the localization length exponent 
$\tilde\nu = 2.35 \pm 0.07$.\footnote{We have also performed similar 
simulations in the limit when $t_0=1$. As these simulations are {\it much\/} 
less computer intensive, we have averaged up to 100 000 samples of 
systems ranging in size between $L=10$ and $L=960$, obtaining 
$\nu = 1.333$, in accordance with the expected value $4/3$ --- see
Sec.\ 5.2.}

So far, we have only discussed one single critical exponent associated with
the local\-iza\-tion-de\-local\-iz\-ation 
exponent in the quantum Hall effect, namely
$\tilde\nu$, the localization length exponent.  Recent analysis of the
Chalker-Coddington model by Klesse and Metzler [54], which is based on 
earlier work of Pook and Jan{\ss}en [55], shows that the wave function
is {\it multifractal\/} in this model. In terms of
an open system, where $z$ is to be interpreted as a current amplitude, 
the problem   
is operationally analogous to the random resistor network at the percolation
threshold, where it also has been established that the current distribution
is multifractal [56,57].  
\begin{figure}[t]
\vspace*{14cm}
\includegraphics{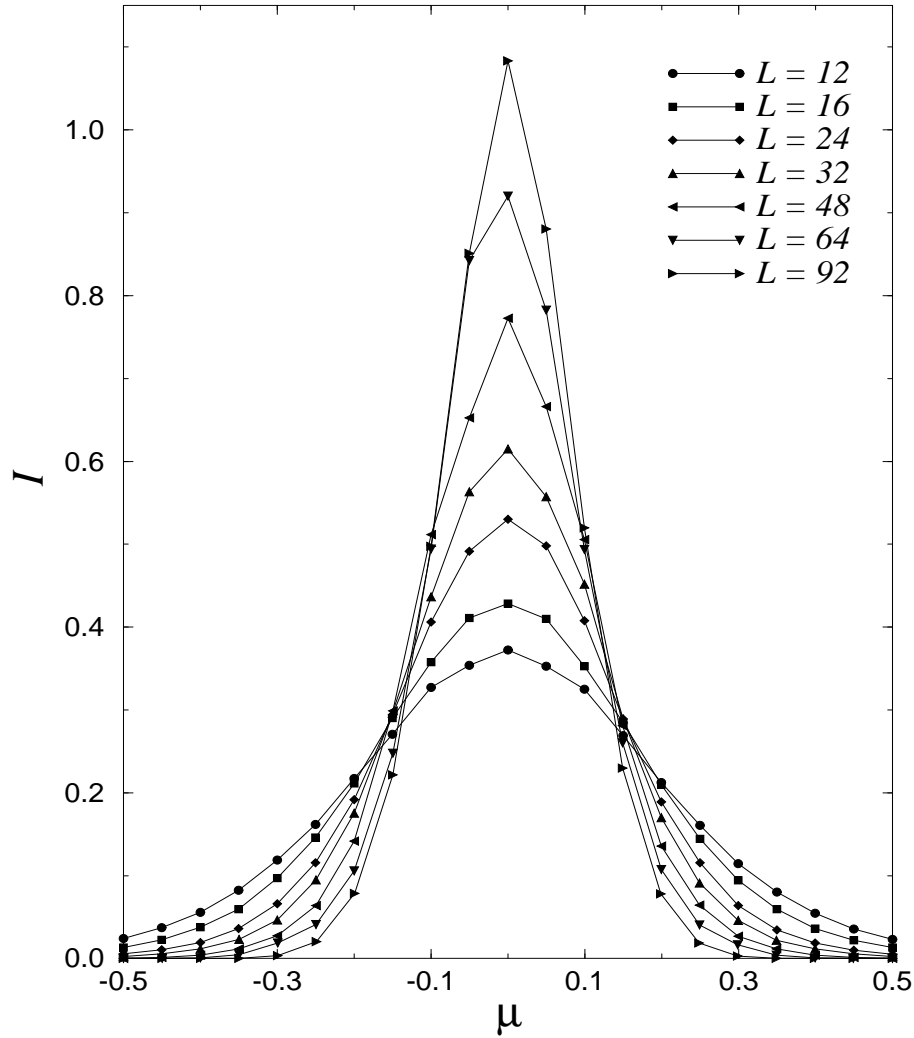}
\caption{The current passing through the system as a
function of chemical potential, for various system sizes.}
\end{figure}

Multifractality means the following: At the critical point, the 
logarithmically binned histogram, $H$, 
of currents $z$ show the following scaling 
form:
$$
H(z,L) \sim L^{f(\alpha)}\;,
\eqno(122)
$$
where
$$
z \sim L^{-\alpha}\;,
\eqno(123)
$$
and $L$ is the lattice size (assuming a square lattice). The two
functions $f$ and $\alpha$ are independent of $L$. If $f(\alpha)$ is a more 
complicated function of $\alpha$ than of linear form $a+b\alpha$, then the 
current distribution is multifractal.
This signifies that there is a continuum of singularities governing the
distribution --- singularities in the sense of non-trivial power laws, 
and not just a small number of exponents, which is usually the case in
connection with critical phenomena.  The geometrical scaling properties of 
percolation, for example, are completely described by two 
exponents, while the current distribution is 
multifractal.\footnote{To clarify Eq.\ (122) and (123) somewhat, 
think of a $d$-dimensional resistor network without percolation disorder:
All bonds are present.  Then, $\alpha=2-d$ and $f=d$. The $f$ {\it vs.\/}
$\alpha$ curve is in this case just a point.}
\begin{figure}[t]
\vspace*{14cm}
\includegraphics{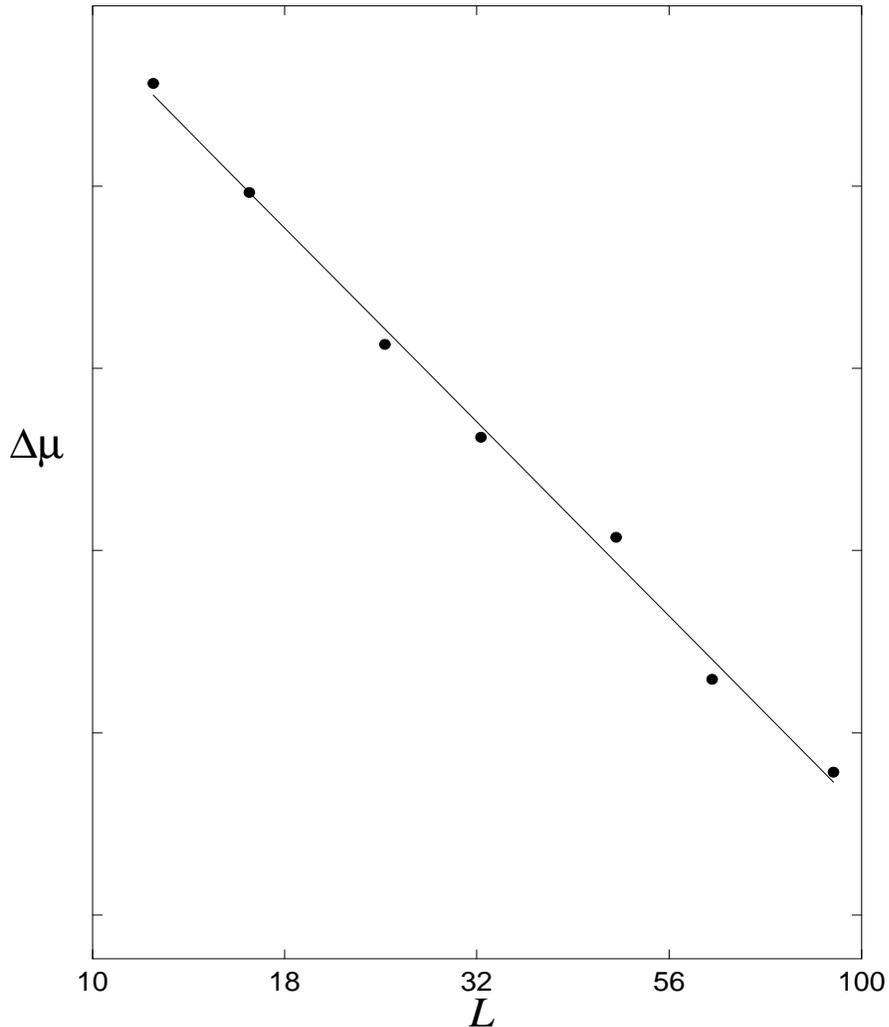}
\caption{Log-log plot of 
the half width of the curves shown in Fig.\ 9. The 
straight line is the result of a least squares fit giving 
$\tilde\nu = 2.35\pm 0.07$.}
\end{figure}

One can study a distribution by building a histrogram.  Equivalently, one
can explore it by measuring its moments. In our case, we define the $n$th
current moment as
$$
M_n=\langle \sum_i |z|^{2n}\rangle\sim L^{y(n)}\;.
\eqno(124)
$$
They all turn out to scale as power laws.  The tell-tale sign of 
multifractality is that
$$
y(n)\neq y(0)+c n\;,
\eqno(125)
$$
where $c$ is a constant. The fundamental relation between $f$, $\alpha$, $y$
and $n$ is of the Legendre type,
$$
y(n)=f-n\alpha\;.
\eqno(126)
$$
This interpretation and equation form the heart of the celebrated $f-\alpha$ 
formalism [58].

We will not pursue this line any further here, except for pointing out that 
this is today an active field of research.
\begin{figure}[t]
\vspace*{12.5cm}
\includegraphics{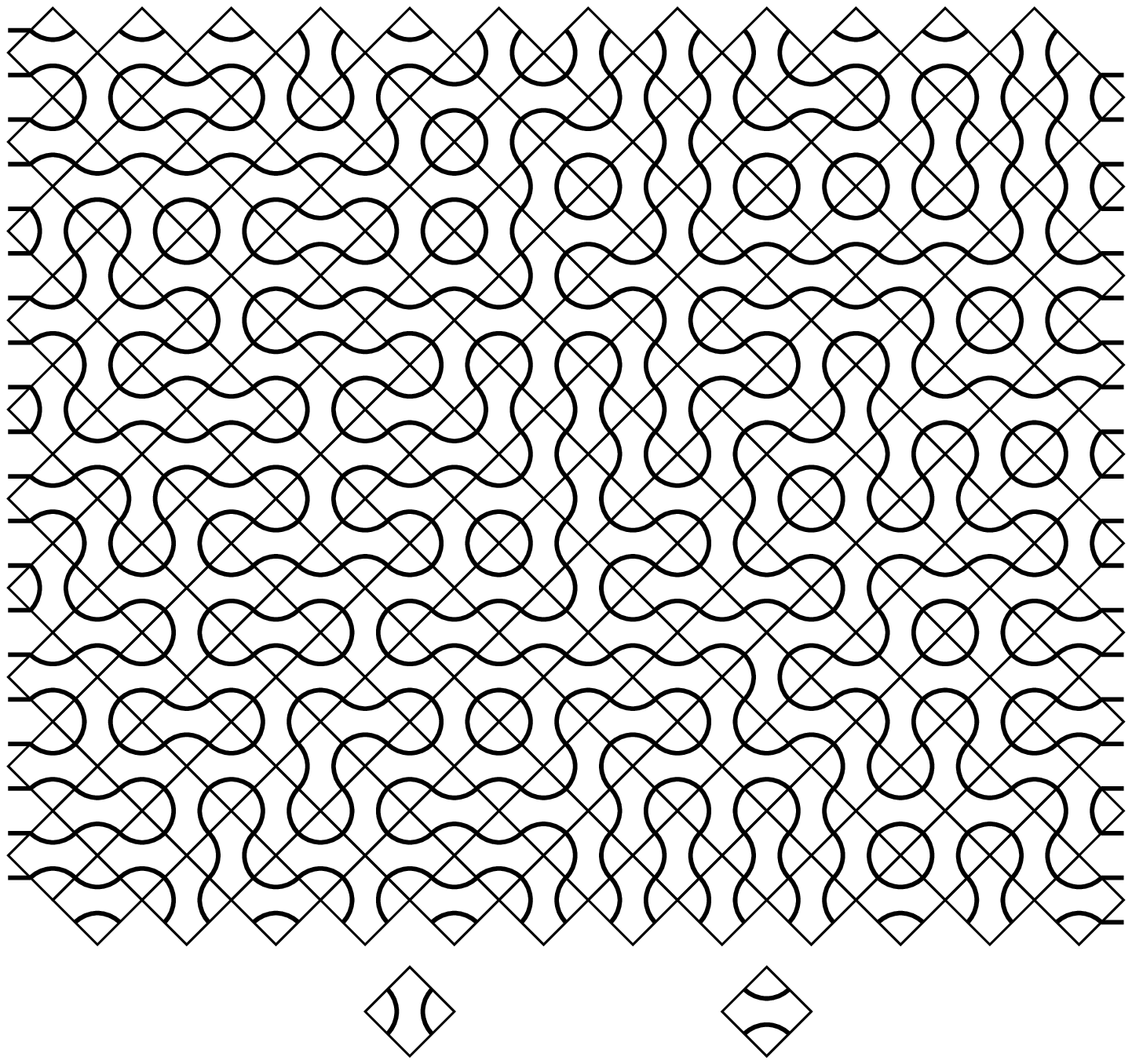}
\caption{The Chalker-Coddingon model in the limit $w\to\infty$
generates ``dragon curves" by randomly tiling the plane with the two
{\it Truchet\/} tiles shown.}
\end{figure}

\bigskip
\centerline{\bf 5.2 Classical Limit of the Chalker-Coddington Model}
\bigskip
The Chalker-Coddington model has an interesting classical limit.
For finite $\mu$ and $v\to\infty$, defined in Eq.\  (110), the scattering 
channels are  $z_1\to z_4$
and $z_2\to z_3$, and when $v\to -\infty$ they are $z_1\to z_3$ and 
$z_2\to z_4$.  In this extreme case, there is no contact across the saddle 
points.  However, delocalization is still possible.  The localization length 
exponent in the extreme case when $w\to\infty$ has been measured by 
Lee {\it et al.\/} [43] to be $1.29\pm0.05$ in agreement with classical 
percolation theory (predicting 4/3).  However, Bratberg {\it et al.\/} [51] 
have noted  that the 
system in this limit is not equivalent to the percolation picture presented
in Sec.\ 4, but to the system shown in Fig.\ 11.
This is a random tiling of the two elements shown at the bottom of the 
figure.  The curves that the markings on the tilings --- known as 
Truchet tiles, see e.g.\ Gale [59] --- produce are known as 
{\it Dragon curves,\/} see Wells [60].  The critical properties of this 
tiling was studied some time ago by Roux {\it et al.\/} [61].  Their 
conclusion was that it is in the universality class of smart kinetic walks 
(SKW).  These may be mapped on the external perimeters of percolation clusters 
in two dimensions. As a consequence, their correlation length exponent 
is the same as in percolation: $\nu_{SKW}=\nu=4/3$. Thus, the 
Chalker-Coddington model with its good exponent $\tilde\nu=2.35\pm 0.03$,
does {\it not\/} simulate the percolation picture of the 
localization-delocalization transition in the integer quantum Hall system
in the ``classical" limit, $w\to \infty$, where $w$ is the width of the
distribution from which $v$ is chosen. The criticality of the model in 
this limit comes from the loop size distribution, and not from joining of 
overlapping loops at saddle points, as there are no saddle point crossings.

\bigskip
\centerline{\bf 6.\ Conclusion}
\bigskip

We summarize in a few sentences the picture of the integer quantum Hall
effect that we have presented here.  Non-interacting electrons move 
two-dimensionally in a strong perpendicular magnetic field and in a 
disordered potential.  Classically, the perpendicular magnetic field causes
the electrons to move in circles about the so-called ``guiding centers."
Quantum mechanically, this circular motion is quantized into Landau levels.  
We assume the disorder potential to be slowly varying on the scale of the
radii of the Landau levels, so that there are no transitions between these
caused by the potential.  The Schr{\"o}dinger Eq.\ (45) then simplifies to 
Eq.\ (49), which shows that the energetically allowed regions where the
electron may move are the equipotential curves of the disordered potential.
The wave function is localized to bands of width of the order of the radius 
of the Landau level, and follows the equipotential curves of the 
potential.  By adjusting the perpendicular magnetic field, these curves
move up or down in the disordered potential landscape.  At a particular 
critical value of the magnetic field, there will be a continuous path of
equipotential curves that span the system.  The electrons in this Landau
level are then delocalized, and there is longitudinal conductance.  For
other values of the magnetic field, they are not.  This mechanism is
repeated for every filled Landau level up to the Fermi level.  The 
localization-delocalization transition is driven by coalescence of 
disjoint equipotential curves, and strongly resembles classical percolation.
However, tunneling at the saddle points and interference effects change
the universality class of the transition from that of percolation.  There
are semi-classical arguments that reproduce the delocalization length 
exponent seen numerically and experimentally, which essentially describe
the localization-delocalization transition as a ``dressed" percolation
process. However, these arguments have serious weaknesses, and it seems
likely that the transition is a pure quantum one. 

The number of experiments that directly test the scaling properties of
the transition are few and to a certain degree not consistent with each 
other.  Why do for instance Dolgopolov {\it et al.\/} [35] find a 
localization exponent $\tilde\nu \approx 1$ which is roughly 
consistent with the {\it classical\/} value 
$\nu=4/3$, rather than the quantum value $\approx 2.3$?  The argument
used to derive the exponent $\tilde\nu\approx 1$ in the work of
Dolgopolov {\it et al.\/} [34] is equivalent
to the gradient percolation argument of Sapoval {\it et al.\/} [62], which 
is in fact {\it incompatible\/} with the slowly-varying landscape model 
presented in these lectures.  

This is clearly not a closed subject.

\bigskip
\centerline{\bf 7.\ Acknowledgements}
\bigskip
We thank I.\ Bratberg, J.\ Hajdu, J.\ S.\ H{\o}ye, B.\ Huckestein, 
M.\ Jan{\ss}en, J.\ Kert{\'e}sz, R.\ Klesse, J.\ M.\ Leinaas, 
C.\ A.\ L{\"u}tken, M.\ Metzler, J.\ Myrheim, K.\ Olaussen, J.\ Piasecki 
and D.\ Polyakov for discussions. This work was in part supported by the 
Norwegian Research Council.  A.H.\ furthermore thanks J.\ Hajdu and SFB 341 
for an invitation to Cologne where part of this review was written.

\bigskip
\centerline{\bf References}
\bigskip
\begin{description}
\item{\hspace*{6pt}[1]} {P.\ W.\ Anderson, Phys.\ Rev.\ {\bf 109}, 1492 (1958).}

\item{\hspace*{6pt}[2]} P.\ W.\ Anderson in {\sl Les Prix Nobel 1977\/} (Almqvist and 
Wiksell, Stockholm, 1978).

\item{\hspace*{6pt}[3]} P.\ J.\ Flory, J.\ Am.\ Chem.\ Soc.\ {\bf 63}, 3083 (1941);
{\it ibid\/} {\bf 63}, 3091 (1941).

\item{\hspace*{6pt}[4]} A.\ S.\ Skal and B.\ I.\ Shklovskii, Sov.\ Phys.\ Semicond.\
{\bf 8}, 1029 (1974).

\item{\hspace*{6pt}[5]} M.\ Sahimi in {\sl Annual Review of Computational Physics,\/}
Vol.\ II, ed.\ by D.\ Stauffer (World Scientific, Singapore, 1995).

\item{\hspace*{6pt}[6]} D.\ J.\ Thouless in {\sl Ill-condensed Matter,\/} ed.\ by R.\ 
Balian, R. Maynard and G.\ Toulouse (North-Holland, Amsterdam, 1979).

\item{\hspace*{6pt}[7]} N.\ W.\ Ashcroft and N.\ D.\ Mermin, {\sl Solid State Physics\/}
(Saunders Collegem Philadelphia, 1976).

\item{\hspace*{6pt}[8]} K.\ von Klitzing, G.\ Dorda and M.\ Pepper, Phys.\ Rev.\ Lett.\
{\bf 45}, 494 (1980).

\item{\hspace*{6pt}[9]} M.\ A.\ Paalanen, D.\ C.\ Tsui and A.\ C.\ Gossard, Phys.\ Rev.\
B {\bf 25}, 5566 (1982).

\item{[10]} R.\ E.\ Prange and S.\ M.\ Girvin, {\sl The Quantum Hall Effect\/}
(Springer Verlag, Berlin, 1987).

\item{[11]} M.\ Jan{\ss}en, O.\ Vieweger, U.\ Fastenrath and J.\ Hajdu, 
{\sl Introduction to the theory of the integer quantum Hall effect\/}
(VCH, Weinheim, 1994).

\item{[12]} T.\ Chakraborty and P.\ Pietil{\"a}inen,
{\sl The quantum Hall effects\/} (2nd ed., Springer, Heidelberg, 1995).

\item{[13]} B.\ Huckestein, Rev.\ Mod.\ Phys.\ {\bf 67}, 357 (1995).

\item{[14]} R.\ Kubo, S.\ J.\ Miyake, N.\ Hashitsume, Solid State Physics 
{\bf 17}, 269 (1965).

\item{[15]} M.\ Tsukada, J.\ Phys.\ Soc.\ Japan, {\bf 41}, 1466 (1976).

\item{[16]} W.\ Gr{\"o}bli, Vierteljahrschrift der Naturforschenden 
Gesellschaft in Z{\"u}rich, {\bf 22}, 37, (1877); {\it ibid\/}
{\bf 22}, 129 (1877).

\item{[17]} J.\ L.\ Synge, Can.\ J.\ Phys.\ {\bf 26}, 257 (1948).

\item{[18]} H.\ Aref, Phys. Fluids {\bf 22}, 393 (1979).

\item{[19]} H.\ A.\ Fertig, Phys.\ Rev.\ B {\bf 38}, 996 (1988).

\item{[20]} G.\ V.\ Mil'nikov and I.\ M.\ Sokolov, JETP Letters {\bf 48}, 
536 (1988).

\item{[21]} A.\ Hansen, Springer Lect.\ Notes on Physics, {\bf 437}, 331
(1994).

\item{[22]} D.\ ter Haar, {\sl Selected Problems in Quantum Mechanics\/}
(Infosearch, London, 1964).

\item{[23]} L.\ D.\ Landau and E.\ M.\ Lifshitz, {\sl Quantum Mechanics\/}
(Pergamon, Oxford, 1965).

\item{[24]} M.\ B{\"u}ttiker, Phys.\ Rev.\ Lett.\ {\bf 57}, 1761 (1986).

\item{[25]} B.\ I.\ Halperin, Phys.\ Rev.\ B {\bf 25}, 2185 (1982).

\item{[26]} M.\ B\"uttiker, Phys.\ Rev.\ B {\bf 38}, 9375 (1988).

\item{[27]} R.\ F.\ Kazarinov and S.\ Luryi, Phys.\ Rev.\ B {\bf 25}, 7626 
(1982).

\item{[28]} S.\ A.\ Trugman, Phys.\ Rev.\ B {\bf 27}, 7539 (1983).

\item{[29]} S.\ Koch, R.\ J.\ Haug, K.\ von Klitzing and K.\ Ploog, Phys.\ 
Rev.\ Lett.\ {\bf 67}, 883 (1991).

\item{[30]} S.\ Koch, R.\ J.\ Haug, K.\ von Klitzing and K.\ Ploog, Phys.\ 
Rev.\ B {\bf 46}, 1596 (1992).

\item{[31]} H.\ P.\ Wei, D.\ C.\ Tsui, M.\ A.\ Paalanen and A.\ M.\ M.\ 
Pruisken, Phys.\ Rev.\ Lett.\ {\bf 61}, 1294 (1988).

\item{[32]} H.\ P.\ Wei, L.\ W.\ Engel and D.\ C.\ Tsui, Phys.\ Rev.\ B 
{\bf 50}, 14609 (1994).

\item{[33]} A.\ A.\ Shashkin, V.\ T.\ Dolgopolov and G.\ V.\  Kravchenko, 
Phys.\ Rev.\ B {\bf 49}, 14486 (1994).
         
\item{[34]} A.\ A.\ Shashkin, V.\ T.\ Dolgopolov, G.\ V.\ Kravchenko, M.\ 
Wendel, R.\ Schuster, J.\ P.\ Kotthaus, J.\ R.\ Haug, K.\ von Klitzing, K.\
Ploog, H.\ Nickel and W.\ Schlapp, Phys.\ Rev.\ Lett.\ {\bf 23}, 3141 (1994).

\item{[35]} V.\ T.\ Dolgopolov, A.\ A.\ Shashkin, G.\ V.\ Kravchenko, C.\
J.\ Emeleus and T.\ E.\ Whall, JETP Letters, {\bf 62}, 168 (1995).

\item{[36]} B.\ Mieck, Europhys.\ Lett.\ {\bf 13}, 453 (1990).

\item{[37]} B.\ Huckestein and B.\ Kramer, Phys.\ Rev.\ Lett.\ {\bf 64}, 
1437 (1990).

\item{[38]} Y.\ Huo and R.\ N.\ Bhatt, Phys.\ Rev.\ Lett.\ {\bf 68}, 1375 
(1992).

\item{[39]} B.\ Huckestein, Europhys.\ Lett.\ {\bf 20}, 451 (1992).

\item{[40]} D.\ Liu and S.\ Das Sarma, Phys.\ Rev.\ B {\bf 49}, 2677 (1994).

\item{[41]} P.\ M.\ Gammel and W.\ Brenig, Phys.\ Rev.\ Lett.\ {\bf 73}, 
3286 (1994).

\item{[42]} J.\ T.\ Chalker and P.\ D.\ Coddington, J.\ Phys.\ C {\bf 21}, 
2665 (1988).

\item{[43]} D.\ H.\ Lee, Z.\  Wang and S.\ Kivelson, Phys.\ Rev.\ Lett.\ 
{\bf 70}, 4130 (1993).

\item{[44]} D.\ K.\ K.\ Lee and J.\ T.\ Chalker, Phys.\ Rev.\ Lett.\ 
{\bf 72}, 1510 (1994).

\item{[45]} A. Hansen and C.\ A.\ L{\"u}tken, Phys.\ Rev.\ B {\bf 51}, 5566
(1995).

\item{[46]} H.\ L.\ Zhao and S.\ Feng, Phys.\ Rev.\ Lett.\ {\bf 70}, 4134
(1993).

\item{[47]} A.\ Coniglio, J.\ Phys.\ A {\bf 15}, 3829 (1982).

\item{[48]} A.\ Hansen and J.\ Kert{\'e}sz, preprint, 1997.

\item{[49]} L.\ Jaeger, J.\ Phys.\ C {\bf 3}, 2441 (1991).

\item{[50]} A.\ Mackinnon and B.\ Kramer, Phys.\ Rev.\ Lett.\ {\bf 47}, 
1546 (1981).

\item{[51]} I.\ Bratberg, A.\ Hansen and E.\ H.\ Hauge,
Europhys.\ Lett.\ {\bf 37}, 19 (1997).

\item{[52]} J.\ M.\ Normand, H.\ J.\ Herrmann and M.\ Hajjar,
J.\ Stat.\ Phys.\ {\bf 52}, 441 (1988).

\item{[53]} A.\ Hansen in {\sl Statistical Models for the Fracture of
Disordered Materials,\/} ed.\ by H.\ J.\ Herrmann and S.\ Roux (North
Holland, Amsterdam, 1990).

\item{[54]} R.\ Klesse and M.\ Metzler, Europhys.\ Lett.\ {\bf 32}, 229 
(1995).

\item{[55]} W.\ Pook and M.\ Jan{\ss}en, Z.\ Phys.\ B {\bf 82}, (1991).

\item{[56]} R.\ Rammal, C.\ Tannous, P.\ Breton and A.\ M.\ S.\
Tremblay, Phys.\ Rev.\ Lett.\ {\bf 54}, 1718 (1985).

\item{[57]} L.\ de Arcangelis, S.\ Redner and A.\ Coniglio, Phys.\ Rev.\
B {\bf 31}, 4725 (1985).

\item{[58]} T.\ C.\ Halsey, M.\ H.\ Jensen, L.\ P.\ Kadanoff, I.\
Procaccia and B.\ Shraiman, Phys.\ Rev.\ A {\bf 33}, 1111 (1986).
         
\item{[59]} D.\ Gale, Math.\ Intelligencer {\bf 17}, (3), 48 (1995).

\item{[60]} D.\ Wells, {\sl The Penguin Dictionary of Curious and 
Interesting Geometry.\/} (Penguin, London, 1991).

\item{[61]} S.\ Roux, E.\ Guyon and D.\ Sornette, J.\ Phys.\ A {\bf 21}, 
L475 (1988).

\item{[62]} B.\ Sapoval, M.\ Rosso and  J.\ F.\ Gouyet, J.\ Phys.\ 
(France) {\bf 46}, L149 (1985).
\end{description}


\end{document}